\documentclass[review,authoryear,12pt]{elsarticle}

\usepackage{graphicx}
\graphicspath{ {./Figures/} }
\usepackage[tight,footnotesize, nooneline]{subfigure}

\usepackage{adjustbox} 
\usepackage[hidelinks]{hyperref} 

\usepackage{amssymb}
\usepackage{amsthm}
\usepackage{amsmath}

\usepackage{lineno}
 
\usepackage{multirow}

\journal{arXiv}

\newcommand\copyrighttext{%
   \textcopyright\ 2016. This manuscript version is made available under the CC-BY-NC-ND 4.0 license \url{http://creativecommons.org/licenses/by-nc-nd/4.0/}}

\begin{document}

\begin{frontmatter}



\title{Optical quantification of harmonic acoustic radiation force excitation in a tissue-mimicking phantom}





\author[Affil1]{Visa Suomi \corref{cor1}}
\author[Affil1]{David Edwards}
\author[Affil1]{Robin Cleveland}
\address[Affil1]{Department of Engineering Science, University of Oxford, Parks Road, Oxford, OX1 3PJ, UK}
\cortext[cor1]{Corresponding Author: Visa Suomi, Institute of Biomedical Engineering, Old Road Campus Research Building, University of Oxford, Oxford OX3 7DQ, UK; Email, visa.suomi@eng.ox.ac.uk; Phone, +44 (0) 1865 617660}

\begin{abstract}
Optical tracking was used to characterize acoustic radiation force (ARF) induced displacements in a tissue-mimicking phantom. Amplitude modulated (AM) 3.3 MHz ultrasound was used to induce ARF in the phantom which was embedded with 10 $\mu$m microspheres that were tracked using a microscope objective and high speed camera. For sine and square AM the harmonic components of the fundamental and second and third harmonic frequencies were measured. The displacement amplitudes were found to increase linearly with ARF up to 10 $\mu$m with sine modulation having 19.5\% lower peak-to-peak amplitude values than square modulation. Square modulation produced almost no second harmonic but energy was present in the third harmonic. For the sine modulation energy was present in the second harmonic and low energy in the third harmonic. A finite element model was used to simulate the deformation and was both qualitatively and quantitatively in agreement with the measurements.
\\
\end{abstract}

\begin{keyword}
Acoustic radiation force \sep optical tracking \sep harmonic displacement \sep amplitude modulation \sep tissue-mimicking phantom \sep tissue deformation \sep finite element analysis
\end{keyword}

\end{frontmatter}

\copyrighttext

\pagebreak









\section*{Introduction}

Ultrasound elastography is an emerging technique used in diagnostic clinical applications in medicine to detect contrast in tissue stiffness \citep{parker2011imaging}. Areas of application include: cancer screening, assessing vascular pathology and monitoring high-intensity focused ultrasound (HIFU) therapy. In many implementations tissue displacement is induced by acoustic radiation force (ARF) which is dependent on both ultrasound intensity and the tissue properties. The work reported here is motivated by a desire to better quantify the ultrasound induced displacement in tissue.

The majority of ARF based ultrasound imaging techniques such as acoustic radiation force impulse (ARFI) imaging \citep{nightingale2002acoustic}, harmonic motion imaging (HMI) \citep{maleke2006single}, supersonic shear imaging (SSI) \citep{bercoff2004supersonic} and electromagnetic acoustic (EMA) imaging \citep{zhang2013electromagnetic} rely on quantifying tissue deformation under ultrasound excitation. The magnitude of this deformation is generally tracked using ultrasound based methods, which use time-delays to estimate the dynamic response of tissue using pulse-echo data \citep{pinton2006rapid}. Although these ultrasound methods can track the displacement at a sub-micrometre scale in the axial direction, they have several limitations. First, the displacement estimation accuracy in the lateral direction is significantly lower compared to the axial direction. While it is possible to derive lateral displacement estimate on the basis of axial strain, it has been shown that the variance of the lateral displacement estimate is 40$\times$(F-number)$^{2}$ worse than that of the axial direction \citep{lubinski1996lateral}. Second, real-time tracking of displacements is limited by the interference from the ARF pulse. This is because the echo signal from the excitation pulse has to be sufficiently attenuated before acquiring the pulse-echo tracking data, which prevents estimating tissue deformation during the excitation phase. Third, the depth of the region-of-interest (ROI) creates physical limitations for the temporal resolution due to the sound speed in tissue ($\approx$1540 m/s). For example, in theory a point located at a focal depth of 7.7 cm in tissue would only allow temporal resolution of 10 kHz, which is within the limit for short duration ARFI pulses. In reality however, the temporal resolution would be even lower due to the interfering echo signals, which would have to be attenuated before starting the next pulse-echo acquisition. Furthermore, ultrasonically derived displacement estimates in solid media are also affected from the so-called shearing artefact \citep{mcaleavey2003estimates}, which is caused by the averaging of individual scatterer amplitudes in displacement calculation. This has been shown to lead to underestimation of tissue peak displacement at the focal point \citep{czernuszewicz2013experimental}. Although ultrasound based tracking methods can be easily implemented in a clinical setting by using the same transducer for pushing and tracking the displacement, the limitations mentioned do not favour their usage in accurate characterisation of tissue dynamics.

\citet{bouchard2009superficial} were the first to demonstrate the feasibility of optical tracking method utilising a high-speed camera to measure the dynamics of tissue deformation under ARF excitations. They used a phantom with embedded steel markers and tracked the displacement of these markers under both impulsive and harmonic excitations. The impulsive excitations were conducted using pulse lengths from 0.5 to 50.0 ms and the harmonic excitations using 50 - 300 Hz amplitude modulation. This tracking data was then compared to tracking data acquired using conventional ultrasound based tracking method. They found the mean peak displacements differ from 2.8 to 30.7\% for impulsive pulses and from --4.3 to 25.3\% for harmonic excitations between the two tracking methods. \citet{bouchard2009optical} also did another study  where they used the same tracking technique to characterise the deformation dynamics in a tissue-mimicking phantom. They used short duration impulsive pulses (0.1 - 0.4 ms) and tracked the induced displacements in both axial and lateral direction with frame rates up to 36 kHz. The displacements were tracked using several individual markers located on- and off-axis, and the results were compared to finite element method (FEM) models.

Later, \citet{czernuszewicz2013experimental} used the same optical tracking method to experimentally validate the displacement underestimation in ultrasound based techniques \citep{mcaleavey2003estimates}. For ultrasound excitation and tracking they used a clinical ultrasound scanner with linear ultrasound probe. The probe was used with F-numbers 1.5 and 3.0 to excite tissue-mimicking phantoms whose Young's moduli were 6.6, 19.8 and 30.2 kPa. They found the displacement underestimation to decrease with higher F-number due to larger cross-sectional area which the force is affecting. Similarly, stiffer phantoms exhibited smaller underestimation error due to faster shear wave spreading, which makes the scatterer movement at the focal point more uniform. Therefore, the maximum underestimation error (35\%) was found using the F-number 1.5 and the softest phantom. These experimental results were in accordance with the FEM simulations by \citet{palmeri2006ultrasonic}.

Several studies have characterised mechanical tissue response to impulsive ultrasound excitation using optical \citep{calle2005temporal, bouchard2009optical, bouchard2009superficial, czernuszewicz2013experimental} and ultrasound tracking methods \citep{palmeri2005finite, wang2014imaging}, but there has been little focus on the tissue dynamics under harmonic excitations. \citet{konofagou2003localized} demonstrated the feasibility of HMI with simulations and experimentally by using ultrasound tracking in tissue-mimicking phantoms. In the experimental part they used three different transducer configurations with ultrasound modulation frequencies ranging from 200 to 800 Hz and four agar phantoms with Young's modulus between 7.1 and 94.6 kPa. The effect of stiffness on the harmonic displacement amplitude was determined and the results were compared to simulations. The experimental results showed the oscillation displacements to vary from --300 to 250 $\mu$m whilst in the simulations the estimated harmonic displacement spanned from --800 to 600 $\mu$m. An exponential decrease of the harmonic displacement amplitude with the phantom stiffness was observed both in simulations and experimentally. Later \citet{konofagou2012harmonic} established the correlation between ultrasonically measured harmonic displacement amplitudes and Young's moduli of tissue-mimicking gel phantoms. They also showed the harmonic displacement amplitude to decrease with increasing Young's modulus of the phantom.

Apart from the effect of tissue stiffness on harmonic displacement amplitude, some research has been done about the effect of modulation frequency on oscillation magnitude. \citet{curiel2009vivo} used HMI to monitor HIFU ablation in rabbit muscle and measured the average normalised harmonic displacement amplitudes as a function of modulation frequency. They used modulation frequencies from 50 to 300 Hz and found the displacement amplitude to decrease with modulation frequency within this range. A similar relation was also observed in simulations performed later by \citet{heikkila2010local}. Furthermore, due to the viscoelastic properties of soft tissues, \citet{liu2007viscoelastic} used ultrasound amplitude modulation for viscoelastic characterisation of tissue properties. They studied the changes in harmonic displacement amplitudes using modulation frequencies from 200 to 1300 Hz in two different phantoms. It was shown that the relative displacement amplitude is dependent on the frequency response of the target material which can be related to the corresponding stiffness. In addition to displacement amplitude the relative phase of the displacement was also shown to vary with modulation frequency.

Previous studies have examined the effect of tissue stiffness and modulation frequency on displacement amplitudes, but several aspects are still to be studied: (i) the dynamics of harmonic excitation pulses using square modulation have not been extensively compared to those of sine modulation, (ii) frequency spectrum analysis has not been conducted to characterise the harmonic frequencies and spectral components of displacement curves and (iii) the absolute harmonic displacement amplitudes and phase shifts have not been analysed in terms of both quantified ARF and modulation frequency. Furthermore, ultrasonic tracking methods have mainly been used previously, which has been shown to suffer from limitations in spatial accuracy, temporal resolution and ultrasound excitation interference. Accurate characterisation of tissue deformation under amplitude modulated (AM) ultrasound waveforms would benefit the development of HMI, EMA and other imaging modalities, which rely on ultrasound pushing the tissue periodically.

This paper focuses on characterising tissue dynamics under square and sine AM ultrasound waveforms. The axial displacement curves generated in a tissue-mimicking phantom are analysed both in the time and frequency domain. The time domain characterisation includes quantifying the absolute peak-to-peak displacements of the harmonic displacement curves by varying the intensity and modulation frequency of the ultrasound field. Furthermore, displacement curves are compared to simulated ARF waveforms to determine the phase shifts of the displacement curves. In the frequency domain, individual frequency components of the displacement curves are identified, and square and sine AM displacement curves are examined to quantify the magnitudes of first, second and third harmonic frequencies. Finally, the experimental results are compared to a FEM model using simulated non-linear ultrasound fields.

\subsection*{Acoustic radiation force}

ARF is a phenomenon where travelling ultrasound waves transfer their momentum to the medium of propagation by the processes of absorption and reflection \citep{torr1984acoustic}. The absorption of ultrasound creates a pushing force in the direction of wave propagation \citep{starritt1991forces}, which results in a displacement of the medium in the same direction. In materials such as soft tissues and tissue-mimicking phantoms, where absorption is the dominating effect \citep{lyons1988absorption}, the magnitude of the force per unit volume $F_{V}$ [kg$\cdot$s$^{-2}\cdot$m$^{-2}$] is given by \citep{torr1984acoustic, starritt1991forces, nyborg1965acoustic}:
\begin{equation}
\label{eq:ARF}
F_{V} = \frac{2 \alpha I_{\mathrm{TA}}}{c}
\end{equation}
where $\alpha$ [Np$\cdot$m$^{-1}$] is the absorption coefficient of the medium, $I_{\mathrm{TA}}$ [W$\cdot$m$^{-2}$] is the temporal average intensity of the ultrasound field at given location and $c$ [m$\cdot$s$^{-1}$] is the sound speed in the medium. In ARF based diagnostic imaging applications focused ultrasound fields are typically used to create local tissue displacements, in which case the greatest force is in the region of the transducer focal point.

\section*{Materials and methods}

\subsection*{Tissue-mimicking phantom}

Paraffin gel-wax was used to construct an optically clear tissue-mimicking phantom. Paraffin gel-waxes are transparent compounds of mineral oil and polymer resin, where the ratio of these two elements is typically 95\% mineral oil and 5\% polymer resin \citep{heilman2002transparent}. Gel-waxes are commonly used in candle-making industry due to their burning properties, but their usage has also been suggested as soft tissue-mimicking materials for ultrasound-guided breast biopsy training \citep{vieira2013paraffin}. 

The acoustic  and mechanical properties of the phantom are presented in Table \ref{tab:gelwax_properties}. The acoustic properties were measured using standard methods \citep{ritchie2013attenuation}. Young's modulus and the frequency response of the phantom material were experimentally measured using dynamic mechanical analysis (DMA Q800, TA Instruments, New Castle, DE, US) from 0 to 50 Hz with multiple repetitions. The shear relaxation modulus and relaxation time parameters (viscosity) were fitted to the frequency response using least squares method \citep{kohandel2005frequency}.

\begin{table}[t]
  \centering
  \caption{Acoustic and mechanical properties of gel-wax phantom}
  	\begin{adjustbox}{max width=\textwidth}
	\begin{tabular}{|c|c|c|c|c|c|}
	\hline
	\begin{tabular}{@{}c@{}}Sound \\ speed\end{tabular} & \begin{tabular}{@{}c@{}}Attenuation \\ 	power law\end{tabular} & Density & \begin{tabular}{@{}c@{}}Young's \\ modulus\end{tabular} & 				\begin{tabular}{@{}c@{}}Shear relaxation \\ modulus\end{tabular} & \begin{tabular}{@{}c@{}}Relaxation \\ time\end{tabular} \\  
	\hline
	\rule{0pt}{12pt} 
	1446 m/s & 0.16 dB/cm/MHz$^{1.67}$ & 0.84 g/cm$^3$ & 7.8 kPa & 4.8 kPa & 0.35 ms\\
	\hline
	\end{tabular}
	\end{adjustbox}
  \label{tab:gelwax_properties}
\end{table}

The tissue-mimicking phantom was constructed by melting 30 ml of gel-wax in a glass container until its temperature reached 90 $^\circ$C. After this, three drops of black polystyrene microspheres (diameter 10.23 $\pm$ 0.344 $\mu$m, Polybead, Polysciences, Eppelheim, Germany) were added to act as markers for optical tracking. The liquid compound was then poured into a 100 mm (lateral) $\times$ 20 mm (axial) $\times$ 15 mm (elevation) phantom holder and was left to cool down at room temperature 21 $\pm$ 1 $^\circ$C. The phantom holder was made of clear plastic for efficient light transmission and had removable sides and a top part to allow ultrasound propagation through the phantom in the axial direction.

\subsection*{Experimental setup and parameters}

\begin{figure}[t]
\begin{center}
\includegraphics[height=6cm]{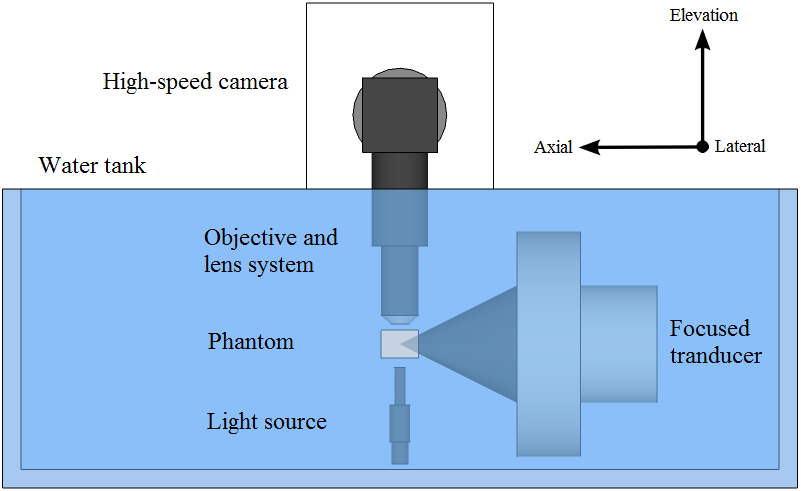}
\end{center}
\caption{Schematic presentation of the experimental setup. The cone coming out of the transducer face illustrates the ultrasound field.}
\label{fig:experimental_setup}
\end{figure}

All the experiments were conducted using the setup shown in Figure \ref{fig:experimental_setup}. A spherically focused single-element transducer (H-102, Sonic Concepts Inc, Bothell, WA, US) was submersed into a water tank which was filled with degassed and distilled water. The transducer had a geometric focal length of 62.6 mm, a 64.0 mm active element diameter with a central hole of 20.0 mm resulting in F-number 0.98, and it was driven at its 3.3 MHz resonance frequency. A needle hydrophone (HMP0075, Precision Acoustics, Dorset, UK) was then attached to a three-axis micro-positioning stage (M-443, Newport Corporation, Irvine, CA, US) and was used to find the focal point of the transducer. The focal point of the ultrasound field was found by manually scanning the hydrophone in axial, elevation and lateral directions. The peak-positive pressure value of the ultrasound waveform was determined in each direction using an oscilloscope (WaveRunner 64Xi-A, Teledyne LeCroy, Spring Valley, NY, US), and the process was repeated until the maximum values of all three axes were found. The needle hydrophone had a physical tip diameter of 300 $\mu$m and an active element diameter of 75 $\mu$m which were smaller than the approximate $-$6 dB beam diameter (BD) of the transducer focal point in water \citep{olympus2011ultrasonic}:
\begin{equation}
\mathrm{BD}_{-6\ \mathrm{dB}} = \frac{1.02lc}{f_{0}d_{\mathrm{a}}} = 450\ \mathrm{\mu m}
\end{equation}
where $l$ [m] is the geometrical focal length of the transducer and $d_{\mathrm{a}}$ [m] is the active element diameter.

After the location of the transducer focal point was determined, the needle hydrophone was left in its position and a mirror system consisting of a 40$\times$ objective (LUMPLFLN 40XW, Olympus Corporation, Essex, UK) was attached to another three-axis micro-positioning stage (M-443, Newport Corporation, Irvine, CA, US). The adjustment in the axial and lateral directions was done so that the tip of the hydrophone needle was in the centre of the image and in the elevation direction the visually sharpest position of the objective was determined. This was done to ensure that the acoustic focal point of the transducer and the optical focal point of the objective were coincident.

After the two focal points were aligned, the needle hydrophone was detached from the positioning stage and the phantom holder was attached instead. The tissue-mimicking phantom was then positioned below the lens system so that the upper surface of the phantom was flush with the end of the objective, and that the ultrasound focal point was approximately in the middle of the phantom in the lateral and axial directions. The objective had a working distance of 3.3 mm, which was the limiting factor for the ultrasound focal point depth in the elevation direction. Below the phantom, a 150-watt halogen cold light source (KL 1500 compact, Schott AG, Mainz, Germany) with an optical light guide (600 mm 1-branch gooseneck, Schott AG, Mainz, Germany) was placed so that the end of the light source was facing the objective, and a gap of approximately 10 mm was left between the phantom holder and the tip of the light source. The light guide was made waterproof by using a heat shrink and a custom made clear plastic top part. 

For data acquisition, a high-speed camera (Memrecam HX-3, NAC Image Technology, Simi Valley, CA, US) was attached to the end of the optical mirror system. The ultrasound transducer was driven using two arbitrary waveform generators (33250A, Agilent Technologies, Santa Clara, CA, US) which were connected in sequence. The first generator was used to create either a 10-cycle square AM waveform with 50\% duty cycle $V_{\textrm{square}}$ [V] or a 10-cycle sine AM waveform as
\begin{equation}
\label{eq:sine_modulation}
V_{\textrm{sine}}=\frac{1}{2}\big(1-\cos(2 \pi f_{\textrm{AM}}t)\big)\sin(2 \pi f_{0}t)
\end{equation}
where $f_{\textrm{AM}}$ [Hz] is the modulation frequency. The second function generator was used to create the carrier wave at 3.3 MHz and to trigger the high-speed camera to start the data acquisition. The AM waveforms were amplified by 55 dB with a RF-amplifier (A300, Electronics \& Innovation, Rochester, NY, US) and applied to the transducer via an impedance matching box. 

The experimental ultrasound parameters measured in distilled/degassed water using a calibrated needle hydrophone (HNA-400, Onda Corporation, Sunnyvale, CA, US) are presented in Table \ref{tab:ultrasound_parameters}. The ultrasound propagation distance in the phantom was short (1 cm) which resulted in negligible reduction of ultrasound pressure ($-$1.86 dB). The frequency response of the hydrophone was taken into account when calculating the ultrasound parameters \citep{howard2007hifu}. Spatial peak temporal average intensity values were calculated using the equation \citep{cobbold2006foundations}:
\begin{equation}
\label{eq:Ispta}
I_{\mathrm{SPTA}} = \frac{1}{t_{\mathrm{pc}}} \int_{0}^{t_{\mathrm{pc}}} \frac{P(t)^{2}}{\rho_{0}c} dt
\end{equation}
where $t_{\mathrm{pc}}$ [s] is the duration of the pressure cycle with the highest peak value and $P(t)$ [Pa] is the corresponding time dependent pressure waveform. Mechanical index (MI) values were calculated using the equation \citep{szabo2004diagnostic}:
\begin{equation}
\label{eq:MI}
\mathrm{MI} = \frac{P_{\mathrm{PN}}}{\sqrt{f_{0}}}
\end{equation}
where $P_{\mathrm{PN}}$ [MPa] is the peak negative pressure of the ultrasound field.

\begin{table}[t]
  \centering
  \caption{Ultrasound parameters used in the experiments for square and sine modulated harmonic excitations}
  	\begin{adjustbox}{max width=\textwidth}
	\begin{tabular}{|r|r|r|r|r|r|r|r|r|r|r|r|}
	\hline
    Experiment no. & 1     & 2     & 3     & 4     & 5     & 6     & 7     & 8     & 9     & 10    & 11 \\
    	\hline
    $f_{\mathrm{AM}}$ (Hz) & \multicolumn{3}{|c|}{50} & \multicolumn{4}{|c|}{160}       & \multicolumn{4}{|c|}{1000} \\
    \hline
    PPP (MPa) & 4.62  & 7.39  & 10.73 & 4.62  & 7.39  & 10.73 & 13.32 & 4.62  & 7.39  & 10.73 & 13.32 \\
    \hline
    PNP (MPa) & $-$3.47 & $-$4.83 & $-$6.25 & $-$3.47 & $-$4.83 & $-$6.25 & $-$7.09 & $-$3.47 & $-$4.83 & $-$6.25 & $-$7.09 \\
    \hline
    $I_{\mathrm{SPTA}}$ (W/cm$^2$) & 534   & 1209  & 2038  & 534   & 1209  & 2038  & 3013  & 534   & 1209  & 2038  & 3013 \\
    \hline
    MI & 1.91  & 2.66  & 3.44  & 1.91  & 2.66  & 3.44  & 3.90  & 1.91  & 2.66  & 3.44  & 3.90 \\
	\hline
	\end{tabular}
	\end{adjustbox}
  \label{tab:ultrasound_parameters}
\end{table}

\subsection*{Data acquisition}

\begin{figure}[t]
    \centering
    \subfigure[]
    {
        \includegraphics[height=6cm]{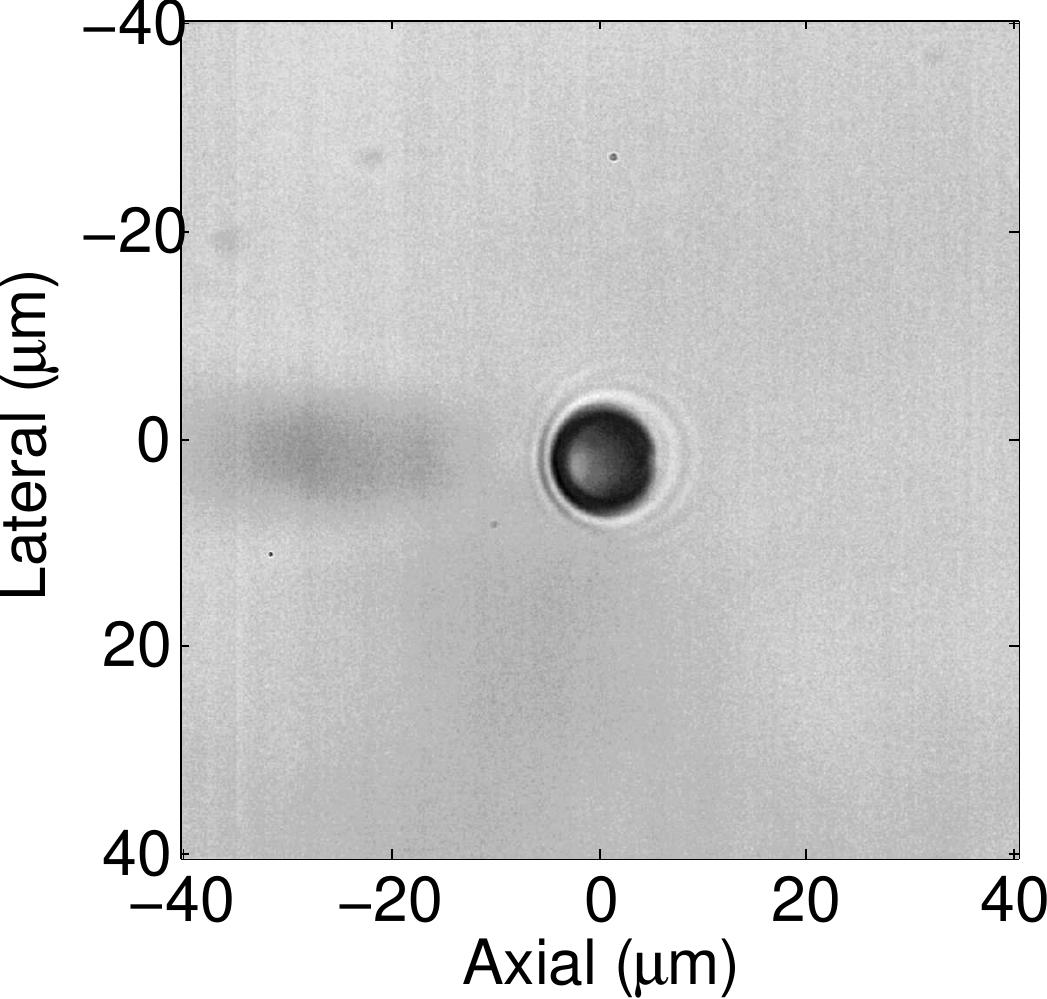}
    }
    \subfigure[]
    {
        \includegraphics[height=6cm]{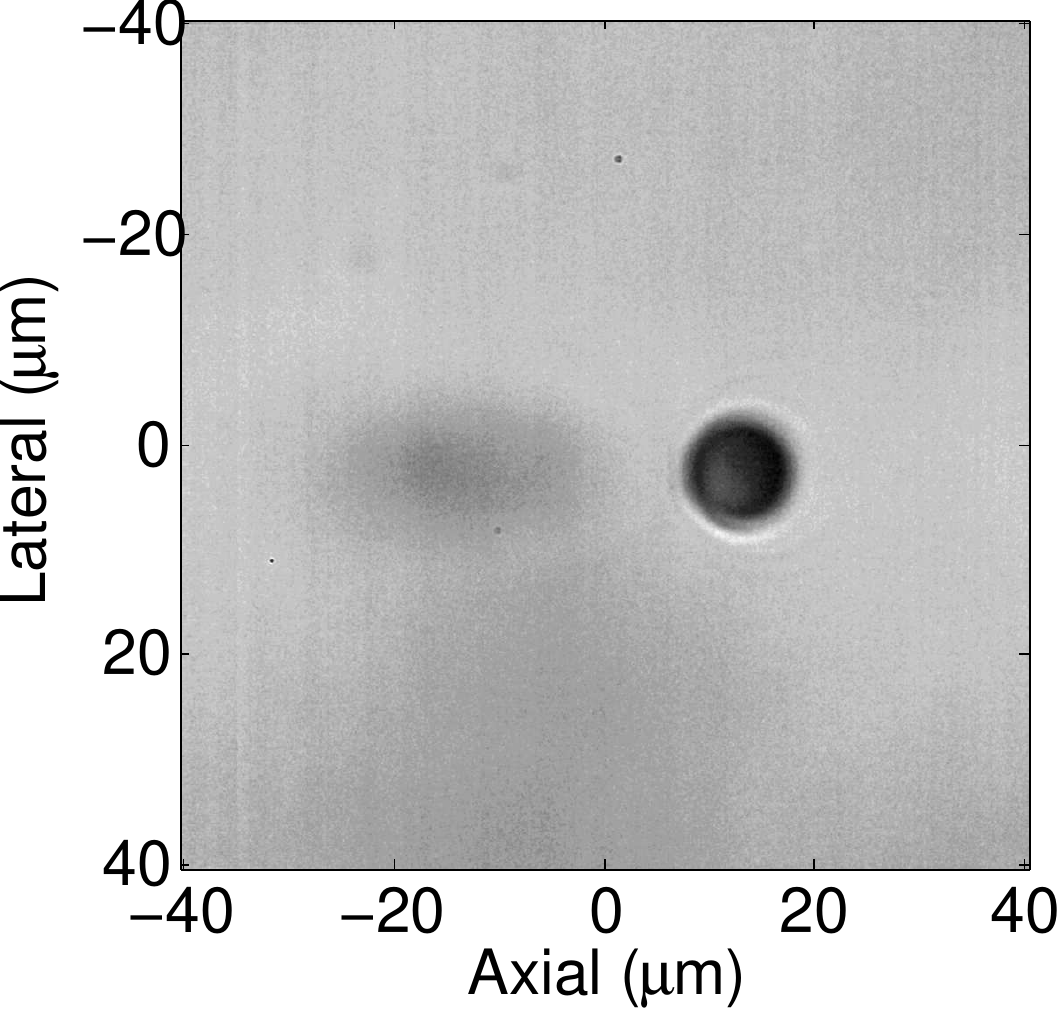}
    }
    \caption{Representative high-speed camera images captured (a) before and (b) during ultrasound excitation. The focal point of the ultrasound field is in the middle of the image.}
    \label{fig:fov}
\end{figure}

A total of 11 experiments listed in Table \ref{tab:ultrasound_parameters} were conducted for both square and sine AM waveforms using the same protocol. An ARF waveform with 10 modulation cycles (parameters in Table \ref{tab:ultrasound_parameters}) was transmitted into the tissue-mimicking phantom through water path. Simultaneously, at the beginning of the ultrasound excitation, a trigger signal was sent to the high-speed camera which started capturing images through the objective and the mirror system. The frame rate of the camera was set to 10 kHz for all the square modulated excitations, and frame rates of 1, 2 and 10 kHz were used for 50, 160 and 1000 Hz sine modulated excitations respectively. This was done to reduce the file size of the acquired image data whilst keeping the frame rate at least 10 times higher than the modulation frequency. The shutter speed of the camera was set to be the inverse of the frame rate. The trigger signal in the camera software was adjusted so that it was in the middle of the image buffer i.e., the same number of frames were captured before and after the ultrasound excitation. Representative image frames recorded before and during ultrasound excitation are shown in Figure \ref{fig:fov}. The field-of-view (FOV) of the objective was kept the same in all the experiments so that the initial position of the microshpere was in the middle of the image. All the measurements were conducted  at room temperature 21 $\pm$ 1 $^\circ$C and the light source was turned off for approximately one minute between the acquisitions to prevent unnecessary heat production in the water bath.

\subsection*{Data analysis}

After each experiment a sequence of images were captured in the camera buffer and a set of images from 5 frames before the beginning of the ultrasound excitation to 100 frames after the end of the excitation were selected. These image sets were then processed through an algorithm pipeline in Matlab R2013b (MathWorks Inc, Natick, MA, US) to calculate axial deformation in time, average peak-to-peak displacements, average phase shifts, frequency spectra and the magnitudes of harmonic frequencies. The image data from all the experiments were analysed using the same method.

The first image of each image set was selected as the reference image where the microsphere was at rest state. The horizontal location of the microsphere central point in the image was manually selected and an intensity threshold value was used to determine the edges of the sphere. By knowing the diameter of the microspheres (10.23 $\mu$m)  the corresponding pixel diameter in the image could be calculated. Each of the images were captured with 1280 (axial) $\times$ 720 (lateral) resolution resulting in a spatial resolution of 0.1 $\mu$m per pixel. After the resolution of the image set was determined, each of the images in the sequence was compared to the reference image using a 2-D cross-correlation algorithm. The maximum value of the cross-correlation output in each image pair determined the axial and lateral displacement of the microsphere, which was saved into a vector variable. Maximum correlation values were typically over 0.8 throughout the whole image set. After all the images in the sequence had been analysed, the displacement vectors were used to plot tissue deformation in the time domain. The data analysis was done based on the axial displacement curves, which are referred to as displacement curves from now on.

\begin{figure}[t]
    \centering
    \subfigure[]
    {
        \includegraphics[width=0.45\textwidth]{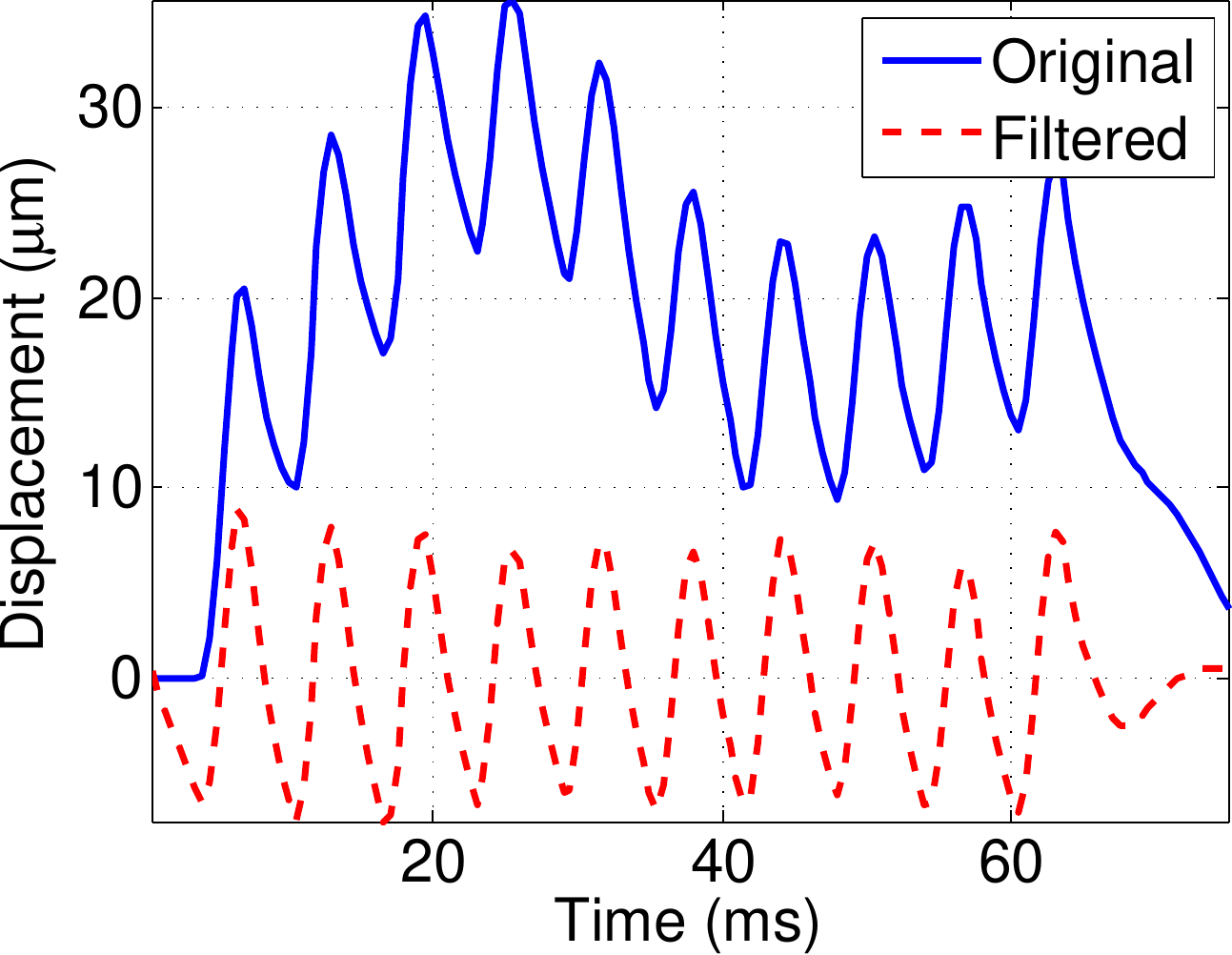}
    }
    \subfigure[]
    {
        \includegraphics[width=0.45\textwidth]{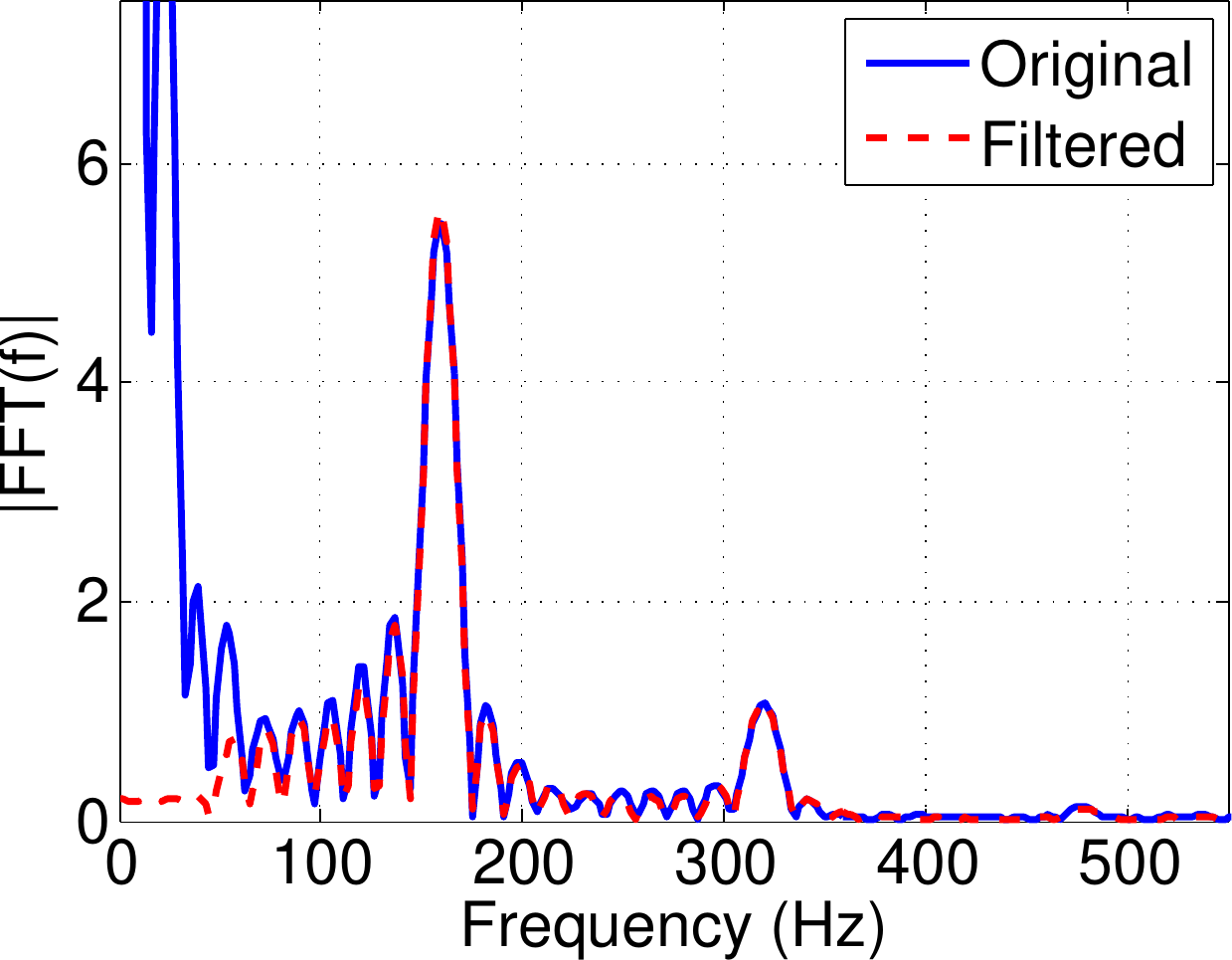}
    }
    \caption{The effect of high-pass filtering on (a) displacement curve and (b) amplitude spectrum when using 160 Hz sine modulation frequency. The data corresponds to the experiment no. 7 in Table \ref{tab:ultrasound_parameters}.}
    \label{fig:example_analysis}
\end{figure}

After obtaining the time domain displacement curves, the average peak-to-peak displacement of each harmonic modulation was calculated. The original waveforms exhibited a baseline drift which was removed by high-pass filtering (see Figure \ref{fig:example_analysis}). High-pass filters with cut-off frequencies of 20, 50 and 250 Hz were used for 50, 160 Hz and 1000 Hz modulation frequencies respectively. After the displacement curves were filtered, the maximum and minimum displacement values of each cycle were determined, and the average peak-to-peak value was calculated based on 8 cycles by ignoring the first and the last cycle of the 10-cycle modulation curve.

The average phase shifts between the displacement curves and the ARF were also determined. This was done by determining the corresponding ARF which was simulated using the mechanical properties of the tissue-mimicking phantom in Table \ref{tab:gelwax_properties} and the experimental ultrasound parameters in Table \ref{tab:ultrasound_parameters}. The magnitude of the phase shift was obtained by comparing the temporal locations of the maximum peak values between the filtered displacement curves and ARF. In the case of square modulation, the middle point of the ARF duty cycle was used as the reference point. Similarly to peak-to-peak displacement, the phase shift in each case was calculated as the mean of eight cycles by ignoring the first and the last cycle of the 10-cycle modulation curve.

The frequency spectra of the displacement curves were calculated using fast Fourier transform (FFT) and the components of the individual spectra were identified. Furthermore, the magnitudes of first, second and third harmonic components were calculated by using discrete Fourier transform (DFT) at each harmonic frequency for displacement curves with different modulation frequencies. In this context first harmonic is used to refer to the fundamental frequency i.e., the ARF modulation frequency.

\subsection*{Simulations}

\begin{figure}[t]
    \centering
    \subfigure[]
    {
        \includegraphics[height=6cm]{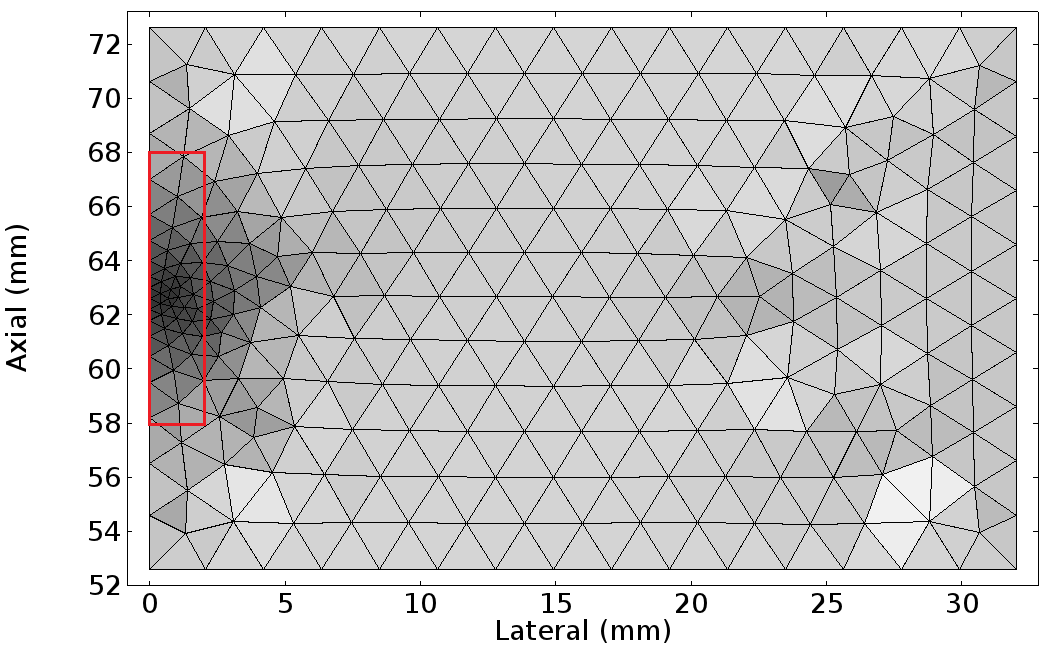}
    }
    \subfigure[]
    {
        \includegraphics[height=6cm]{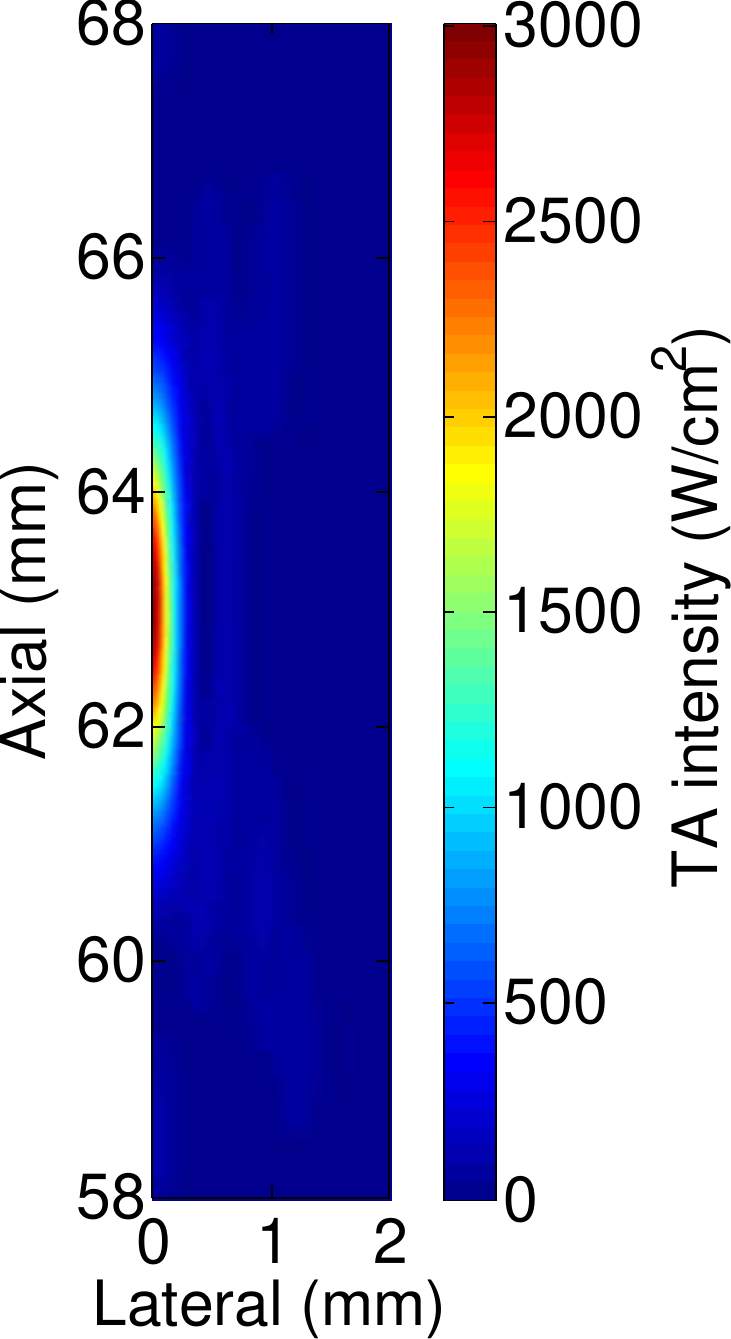}
    }
    \caption{Presentation of (a) the computing mesh and (b) the focal area of simulated ultrasound field (red rectangle) used in the finite element method (FEM) model. Darker gray colours represent smaller elements in the mesh. The ultrasound field simulation corresponds to the experiments no. 7 and 11 in Table \ref{tab:ultrasound_parameters}.}
    \label{fig:simulation_setup}
\end{figure}

The experimental data was compared with the results from a FEM model (Comsol Multiphysics 4.4, Comsol Inc, Burlington, MA, US). The FEM model was constructed using a 2-D axisymmetric phantom with dimensions 32 mm (lateral) $\times$ 20 mm (axial) with the symmetry axis along the axis of the transducer. A free triangular mesh consisting of 478 domain elements and 58 boundary elements was utilised so that the element size in the centre of the phantom, i.e the focal point, was smaller (see Figure \ref{fig:simulation_setup}(a)). A mesh convergence study was performed to determine the element size. The outer surface of the cylinder phantom was constrained while the two surfaces in the axial direction were unconstrained. This mimicked the real situation in the experiments where the phantom was attached to its holder from the both sides in lateral direction. The phantom material was modelled as a linear, isotropic and viscoelastic solid using the experimentally measured parameters in Table \ref{tab:gelwax_properties} and Poisson's ratio $\nu$ = 0.499. The standard linear solid model (Zener) was used for viscoelasticity with the shear relaxation modulus $G$ = 4.8 kPa and relaxation time $\tau$ = 0.35 ms values listed in Table \ref{tab:gelwax_properties}.

The simulations of the acoustic fields were conducted in Matlab R2013b (MathWorks Inc, Natick, MA, US) using an existing HIFU simulator \citep{soneson2009user}. The ultrasound field was calculated as spatial pressure distribution of a spherical transducer by solving the axisymmetric KZK equation \citep{zabolotskaya1969quasi, kuznetsov1971equations} in the frequency domain. The solution takes into account beam diffraction, nonlinear effects, wave reflection at interfaces and the frequency dependence of absorption and phase speed dispersion in the medium. The pressure field of the transducer was solved in the axisymmetric spatial domain of the phantom (i.e., half surface) using 1361 mesh points in the axial and 1696 mesh points in the lateral directions (see Figure \ref{fig:simulation_setup}(b)). The frequency dependent attenuation of the phantom was taken into account by using the experimentally measured parameters in Table \ref{tab:gelwax_properties} and for water typical values at room temperature were employed \citep{pinkerton1949absorption, marczak1997water}. The pressure fields were solved with 128 harmonic frequency components so that the intensity levels matched with the experimentally measured values in Table \ref{tab:ultrasound_parameters}. The corresponding ARF was then calculated using Eq. (\ref{eq:ARF}) for each mesh point taking into account the frequency dependent attenuation.

The simulated ARF mesh was modulated using normalised $P_{\textrm{square}}^2$ or $P_{\textrm{sine}}^2$ and applied as point body loads to the spatial domain of the axisymmetric FEM model. The dynamic response was monitored at the geometric focus of the transducer (62.6 mm). The position of the marker is slightly different from the actual focal point maximum in the phantom (63.2 mm) due to refraction effects. However, this was the case in the experiments as well, where the microsphere was placed in the geometric focus by finding the peak pressure of the focal point in water. The simulated ARF field was modulated using the same modulation frequencies and temporal resolutions as in the experiments.

\section*{Results}

The analysis of the acquired displacement data has been divided into two sections: the time and frequency domains. The time domain section includes graphs of time domain displacement curves, average peak-to-peak displacements and phase shifts for different modulation frequencies and ultrasound intensities. The frequency domain section characterises the frequency spectra and the harmonic frequency components of both sine and square AM displacement curves. In addition to the experimental results, the simulated displacement data is presented and compared to the results in its own section. The possible sources of error are discussed in the last section.

\subsection*{Displacement analysis in the time domain}

\begin{figure}[t]
    \centering
    \subfigure[]
    {
        \includegraphics[width=0.45\textwidth]{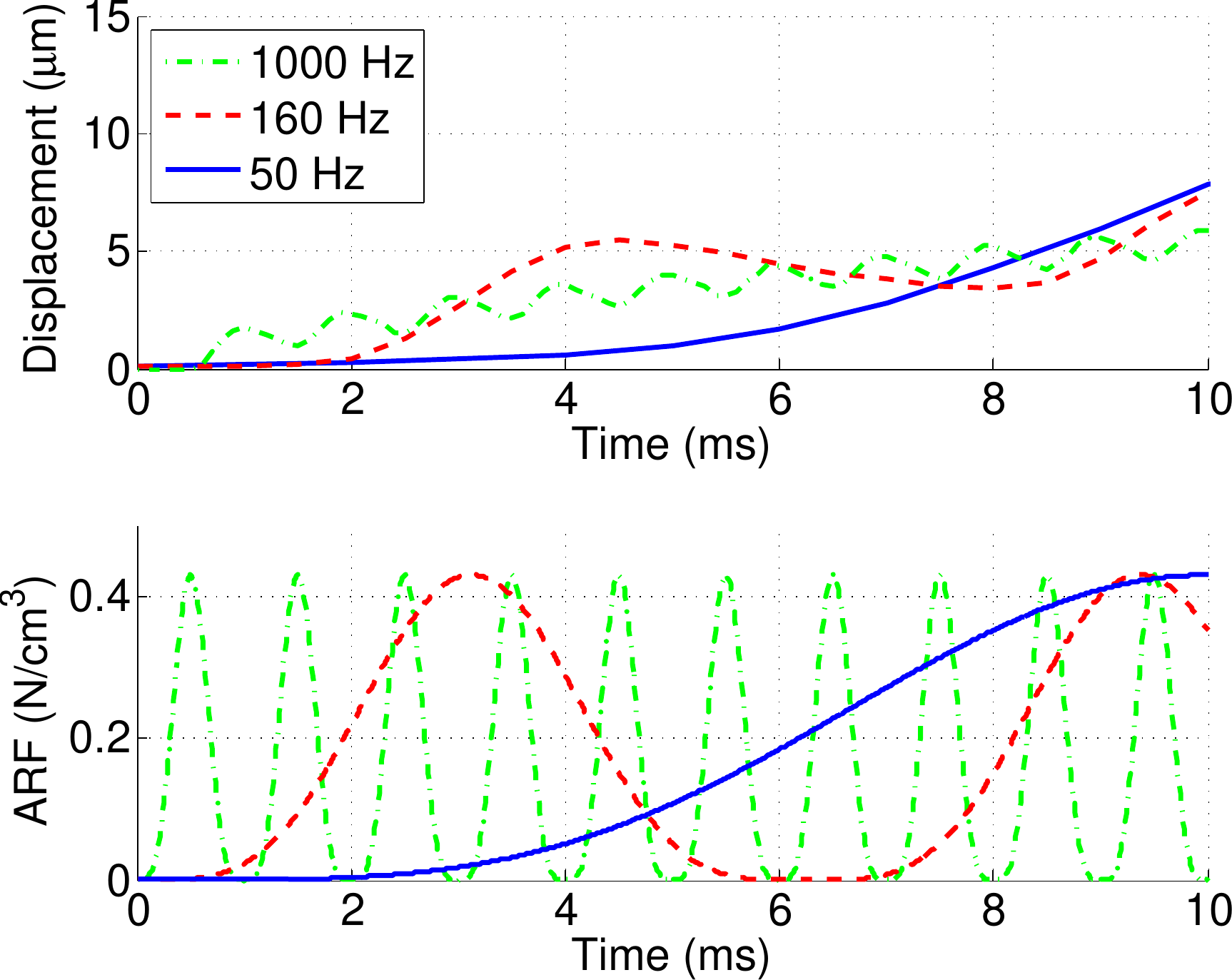}
    }
    \subfigure[]
    {
        \includegraphics[width=0.45\textwidth]{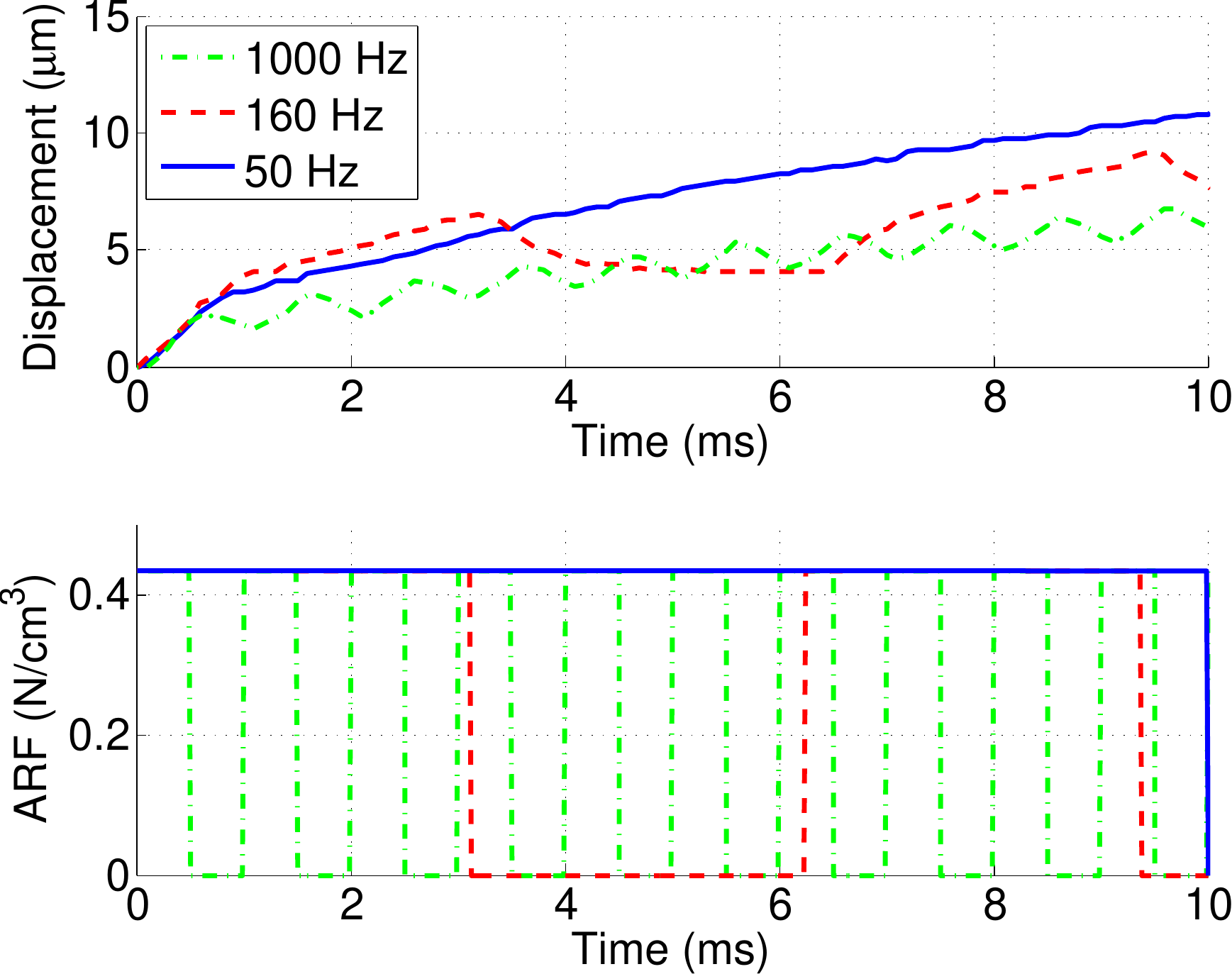}
    }
    \caption{A comparison of (a) sine and (b) square amplitude modulated (AM) waveforms (top) and their corresponding simulated acoustic radiation force (ARF) in the time domain (below). The level of ARF was kept the same and three modulation frequencies of 50, 160 and 100 Hz were used. The first 10 ms seconds are presented in all cases. The data corresponds to the experiments no. 2, 5 and 9 in Table \ref{tab:ultrasound_parameters}.}
    \label{fig:frequency_comparison}
\end{figure}

The measured displacements were recorded so that all ten modulation cycles were captured. The first 10 ms of the obtained displacement curves with the same ultrasound intensity levels (i.e., the experiments no. 2, 5 and 9) are shown in Figure \ref{fig:frequency_comparison} for sine and square AM waveforms as well as the simulated ARF curves in the focal point. All the ARF values shown in the figures are spatial peak values in the phantom. The time axes are scaled so that the arrival of the ultrasound waveform to the phantom (i.e., the onset of ARF) is at 0 ms in all figures.

The finer details in the square AM displacement curves at 50 and 160 Hz are due to the usage of higher frame rates when capturing the image data. Whilst sine modulated data was captured with approximately 10 to 20-fold sampling frequencies compared to modulation frequencies, a constant frame rate of 10 kHz was used for all square AM data. This resulted in finer detail in square modulated displacements at low modulation frequencies, but does not affect the results of the data analysis.

The mean value of ARF is positive with both square and sine AM, which causes the overall upward trend of the displacement curves. This can be seen clearly in Figures \ref{fig:frequency_comparison}(a) and (b), where in the case of 1000 Hz excitation, it has an almost linear slope within the 10 ms duration of the harmonic excitation. When lower modulation frequencies and longer time windows are used, the displacement curve eventually reaches its saturation point and starts to decrease again, recall Figure \ref{fig:example_analysis}(a).

\begin{figure}[t]
    \centering
    \subfigure[]
    {
        \includegraphics[width=0.45\textwidth]{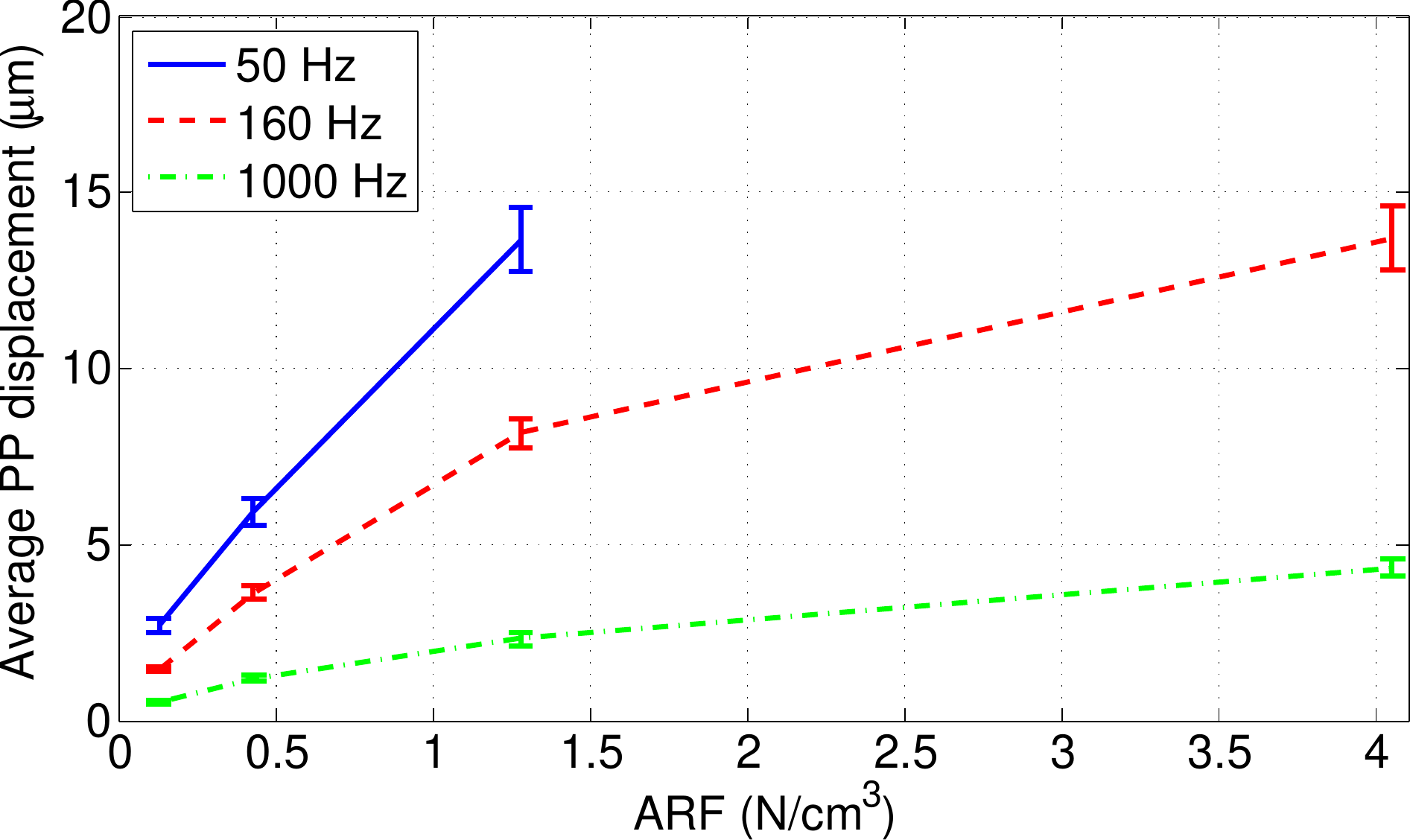}
    }
    \subfigure[]
    {
        \includegraphics[width=0.45\textwidth]{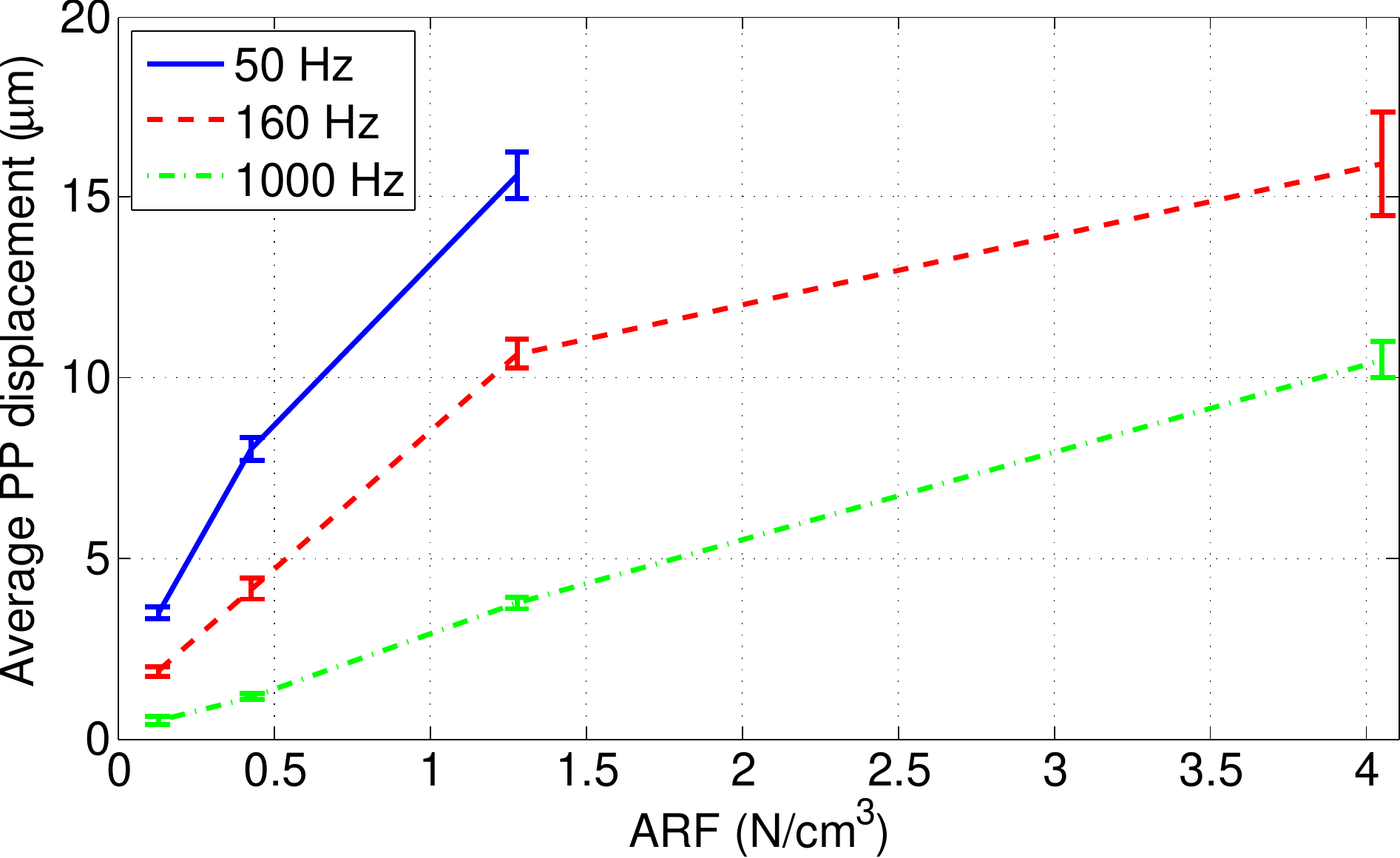}
    }
    \caption{Measured peak-to-peak displacements of (a) sine and (b) square amplitude modulated (AM) waveforms as a function of simulated acoustic radiation force (ARF). Four different ultrasound intensity levels with 50, 160 and 1000 Hz modulation frequencies were used. The magnitude of the displacement is calculated as the mean of eight cycles by ignoring the first and the last cycle of the 10-cycle modulation waveform. The data are expressed as standard deviation of the eight cycles and corresponds to the experiments no. 1-11 in Table \ref{tab:ultrasound_parameters}.}
    \label{fig:arf_displacement}
\end{figure}

The average peak-to-peak displacements as a function of ARF are presented in Figures \ref{fig:arf_displacement}(a) and (b) for sine and square AM waveforms respectively. In each graph three curves are shown representing the three different modulation frequencies used in the experiments. The peak-to-peak displacement amplitudes were calculated as an average of eight cycles after filtering out the low frequency components in the displacement data. The standard deviation within these eight cycles are also shown in each case.

In Figure \ref{fig:arf_displacement} it can be seen that the peak-to-peak displacement increases with applied ARF. The increase appears linear up to a displacement magnitude of about 10 $\mu$m after which the slope decreases. The displacement is higher for lower modulation frequency, which suggests that the medium is stiffer at higher frequencies. This also means the nonlinear effects are more pronounced at low frequencies.

Another observation that can be made from Figure \ref{fig:arf_displacement} is that sine modulated displacement curves have generally lower peak-to-peak amplitude values compared to those of square modulation. On average sine modulated displacement curves had 19.5\% lower peak-to-peak amplitude values, while only in experiments no. 8 and 9 were the sine modulated values 6.7\% and 2.1\% higher compared to square modulation. The minimum average peak-to-peak displacement observed in the results was 0.5 $\mu$m, which was achieved using 1000 Hz modulation frequency and 0.13 N/cm$^{3}$ ARF using both sine and square modulation.

The viscoelastic behaviour of the phantom material could be seen in Figure \ref{fig:frequency_comparison}(a), where the sine modulated displacement curve lags behind the applied ARF. A similar phase shift is also observable in the case of square modulation in Figure \ref{fig:frequency_comparison}(b) when comparing the maximum peak values of the displacement curves to end point of the ARF excitation phase in each cycle.

\begin{figure}[t]
    \centering
    \subfigure[]
    {
        \includegraphics[width=0.45\textwidth]{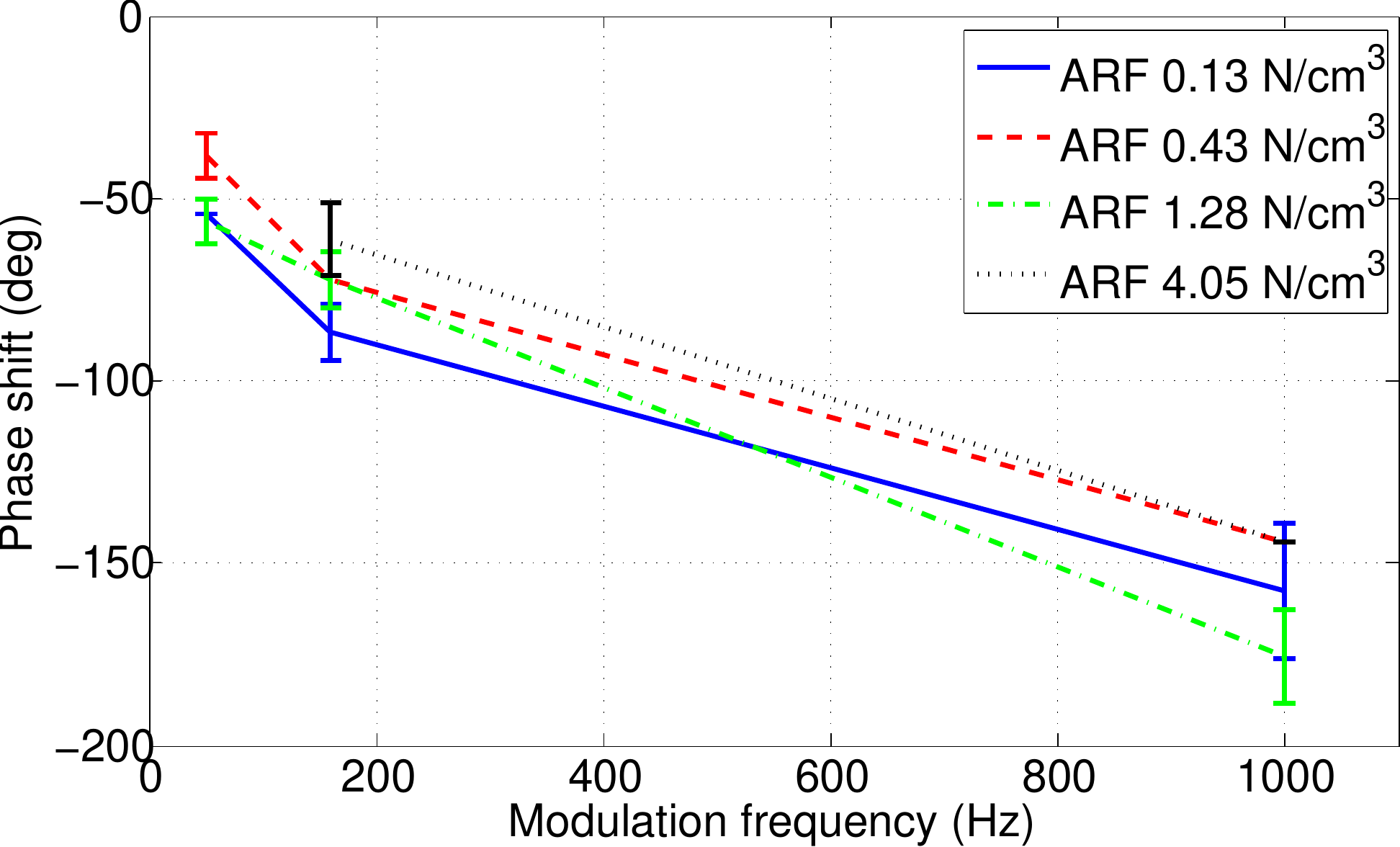}
    }
    \subfigure[]
    {
        \includegraphics[width=0.45\textwidth]{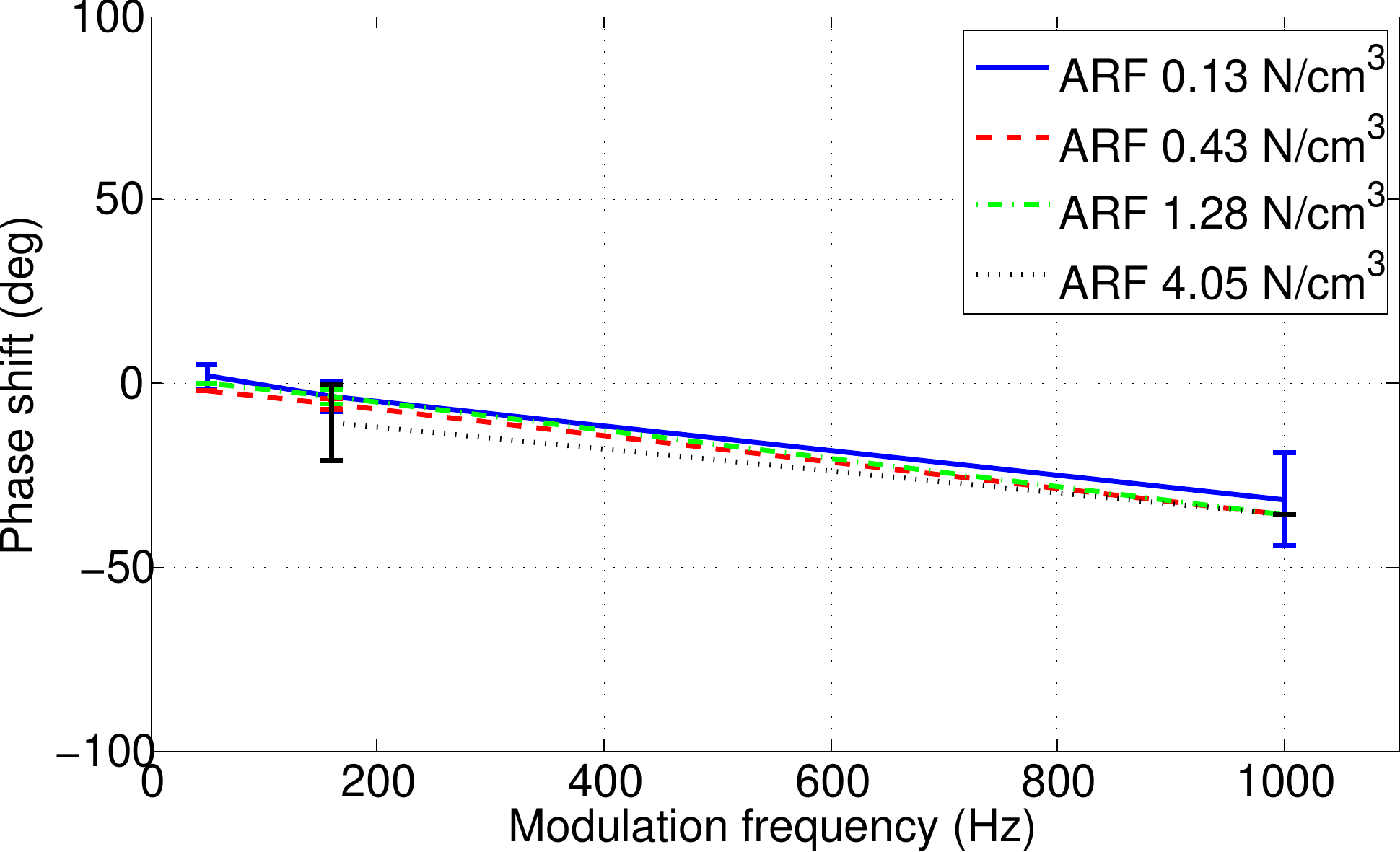}
    }
    \caption{Measured phase shifts relative to acoustic radiation force (ARF) for (a) sine and (b) square amplitude modulated (AM) waveforms as a function of modulation frequency. Negative values represent the lag in the displacement curve compared to the ARF. Four different ultrasound intensity levels with 50, 160 and 1000 Hz modulation frequencies were used. The phase shift was dependent on frequency but not intensity. The data are expressed as standard deviation of the eight cycles and corresponds to the experiments no. 1-11 in Table \ref{tab:ultrasound_parameters}.}
    \label{fig:phase_shift}
\end{figure}

Figure \ref{fig:phase_shift} shows the calculated phase shifts for sine and square modulated displacement curves with respect to the applied ARF. Four different levels of ARF are shown with three measurement points corresponding to the different modulation frequencies. The negative angle value indicates the lag in the displacement curve with respect to ARF. The phase shift values were calculated as an average of eight cycles by ignoring the first and last excitation cycle, and standard deviations are shown within these cycles. It can be seen that the absolute phase shift becomes larger with increasing modulation frequency, but stays approximately the same with different levels of ARF. While the maximum phase shift for sine modulation is $-$175.5$^{\circ}$ at 1000 Hz modulation frequency, the square modulation only reaches $-$36.0$^{\circ}$.

\subsection*{Displacement analysis in the frequency domain}

\begin{figure}[htbp!]
    \centering
    \subfigure[]
    {
        \includegraphics[height=5.5cm]{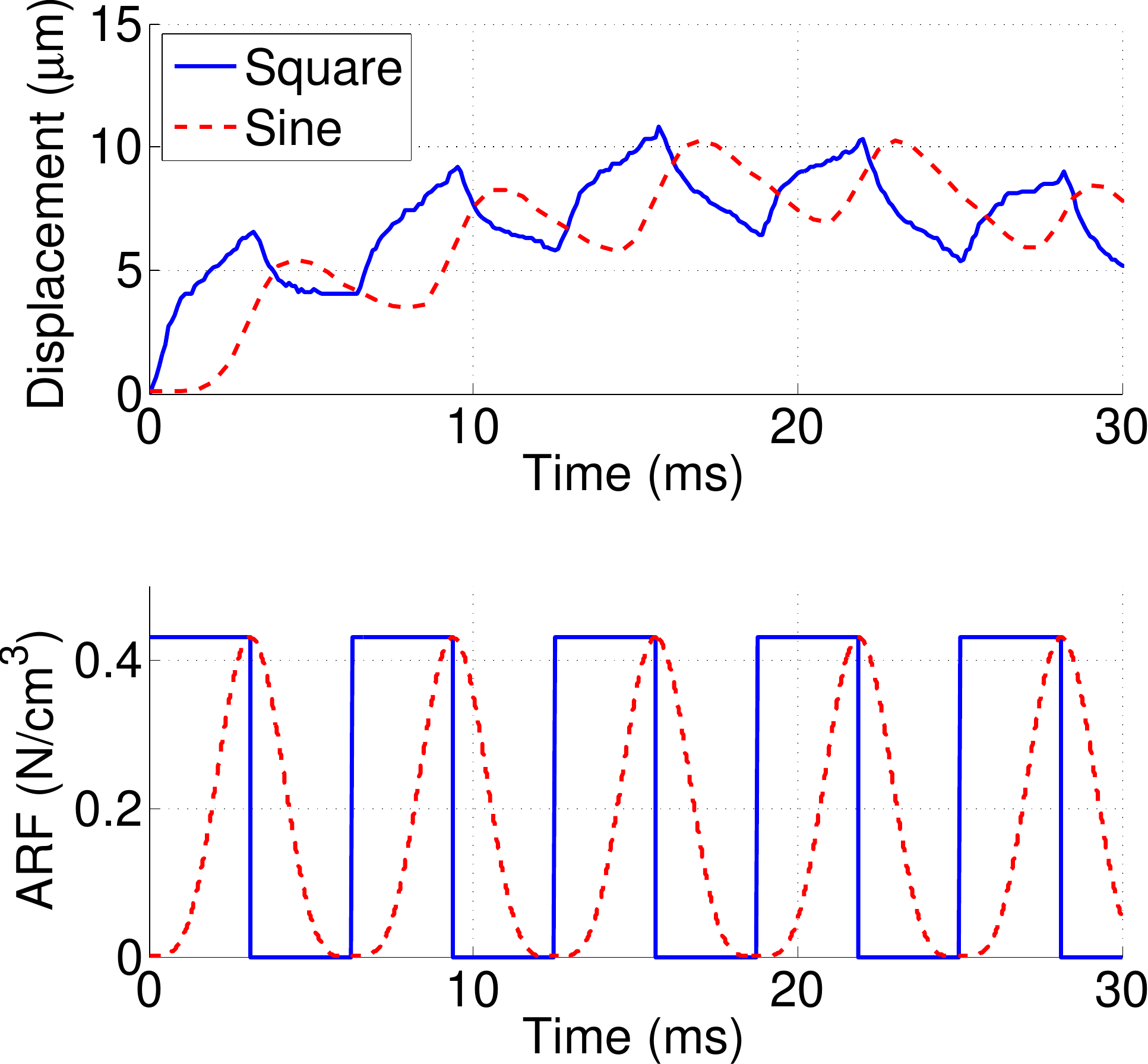}
    }
    \subfigure[]
    {
        \includegraphics[height=5.5cm]{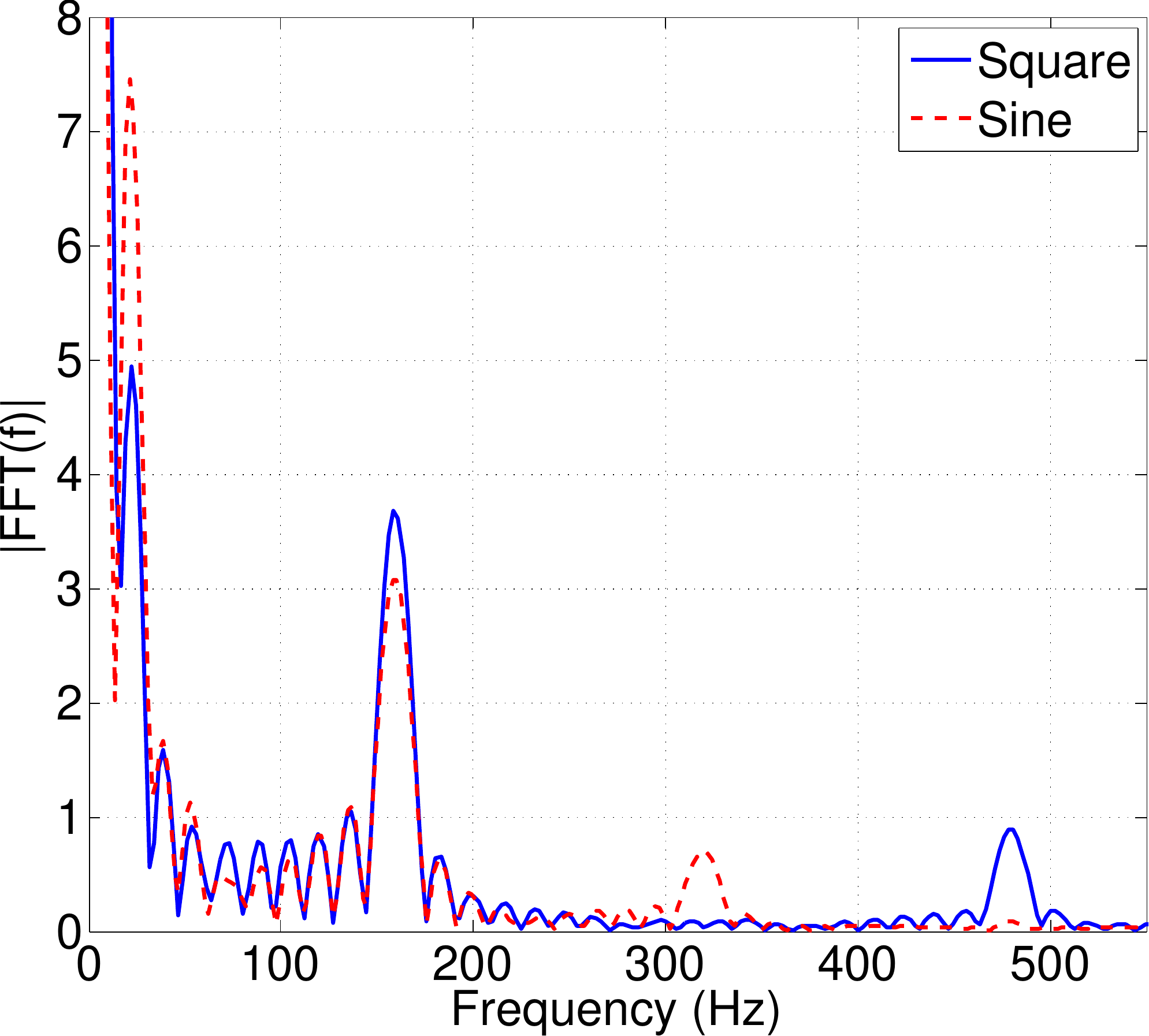}
    }
    \caption{A comparison of 160 Hz square and sine modulated displacement curves (a) in the time domain (above) together with the corresponding simulated acoustic radiation force (ARF) calculations (below) and (b) the frequency spectrum of the both displacement curves. The first 30 ms are presented in the time domain, but the frequency spectrum was calculated based on the full length of the displacement curves. The data corresponds to the experiment no. 5 in Table \ref{tab:ultrasound_parameters}.}
    \label{fig:square_vs_sine_160Hz}
\end{figure}

A comparison of sine and square modulated displacement curves at 160 Hz modulation frequency together with the simulated ARF are presented in Figure \ref{fig:square_vs_sine_160Hz}(a). The first 30 ms are presented in the time domain so that the ultrasound excitation is at the beginning of the curves. Due to the shape of the sine modulated ARF waveform, the resulting displacement curve has an initial offset of about 1.5 ms. Both displacement curves reach their overall peak values at around 15 ms after which they start to decrease again. This behaviour causes lower frequency components in the frequency spectrum as seen in Figure \ref{fig:square_vs_sine_160Hz}(b). This low frequency vibration occurs at around 25 Hz and was observed in the all experiments. The large peak at low frequencies close to 0 Hz (partly cut out of the graph) is caused by the static component in the displacement curves, which causes the harmonic vibration to stay above the 0 $\mu$m rest state through the duration of 10-cycle ARF excitation.

\begin{figure}[htbp!]
    \centering
    \subfigure[]
    {
        \includegraphics[width=0.3\textwidth]{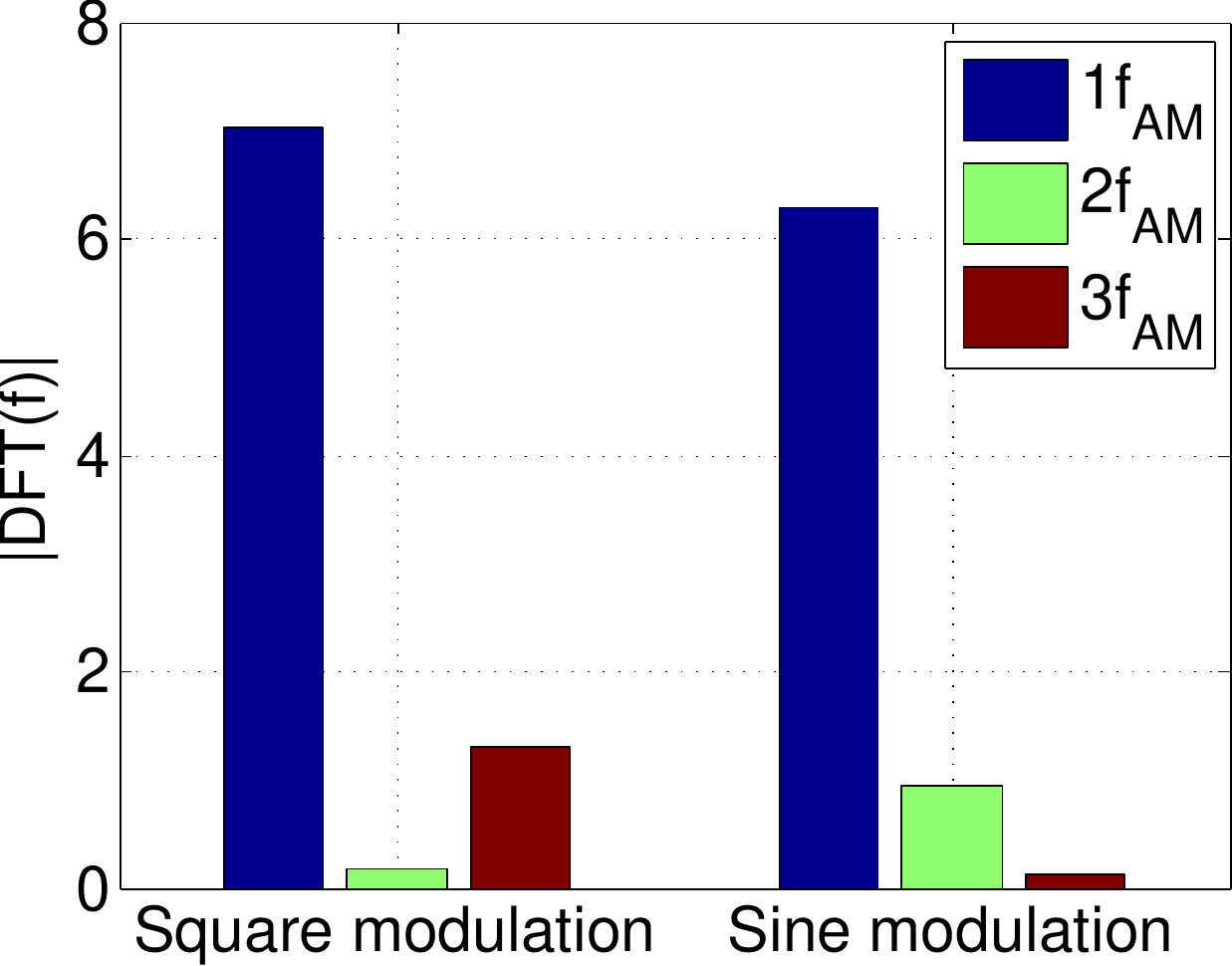}
    }
    \subfigure[]
    {
        \includegraphics[width=0.3\textwidth]{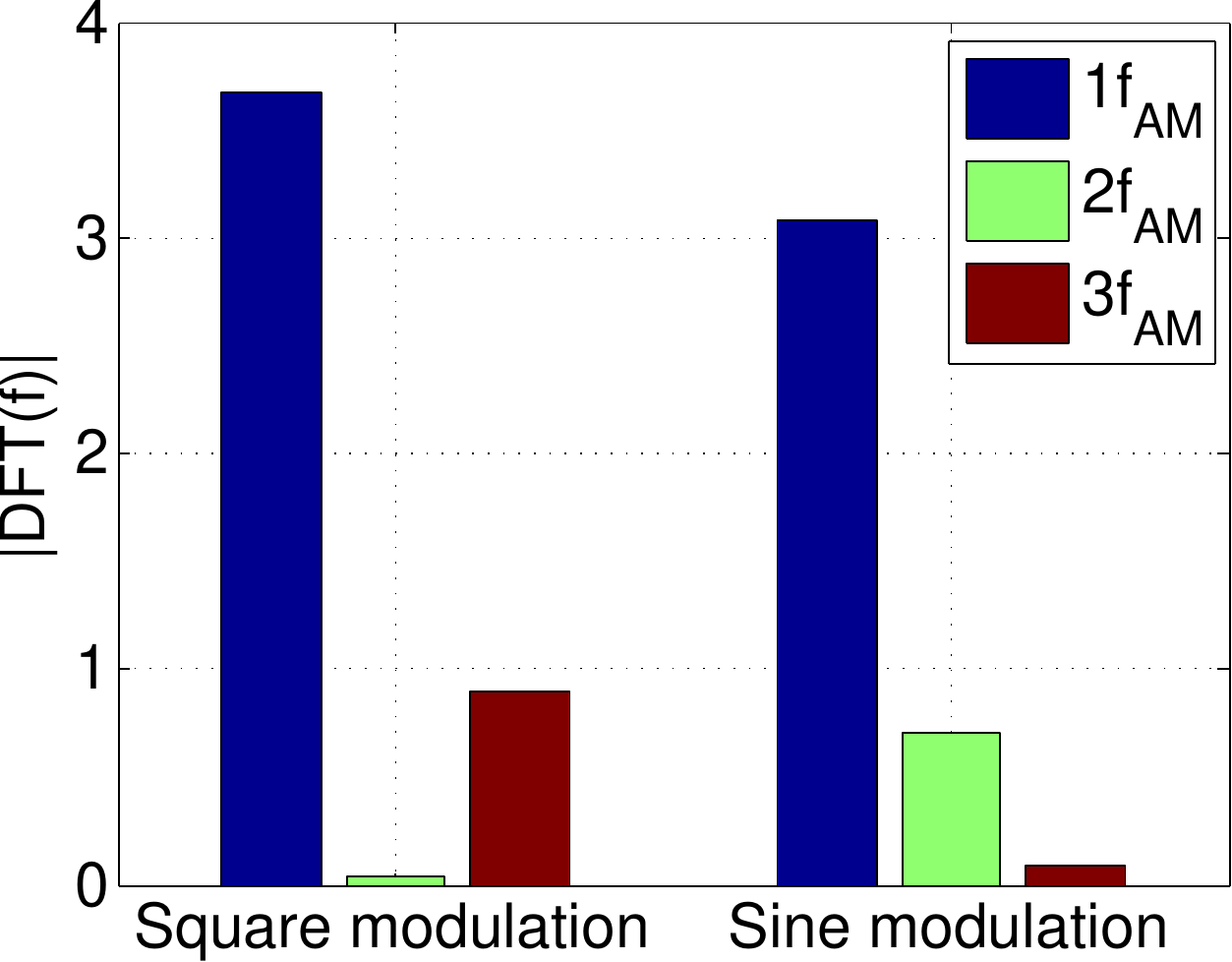}
    }
    \subfigure[]
    {
        \includegraphics[width=0.3\textwidth]{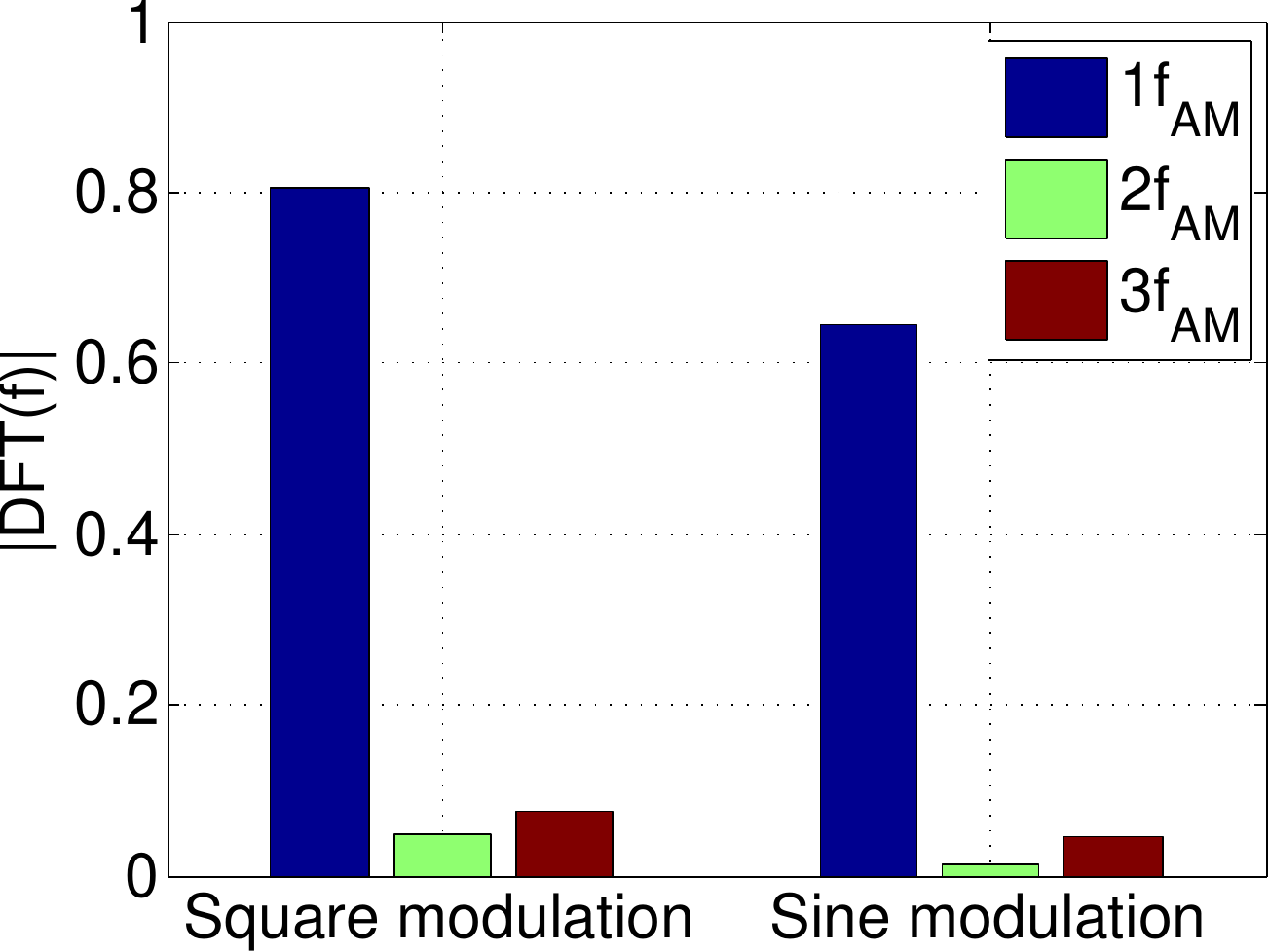}
    }
    \caption{A comparison of the harmonic components of sine and square amplitude modulated (AM) waveforms when using (a) 50, (b) 160 and (c) 1000 Hz modulation frequency. The data corresponds to the experiments no. 3, 6 and 10 in Table \ref{tab:ultrasound_parameters}.}
    \label{fig:DFT_560mVpp}
\end{figure}

The individual frequency components 1$f_{\textrm{AM}}$, 2$f_{\textrm{AM}}$ and 3$f_{\textrm{AM}}$ of displacement curves are presented in Figure \ref{fig:DFT_560mVpp} for 50, 160 and 1000 Hz modulation frequencies used in the experiments no. 3, 6 and 10 respectively. Harmonic component 1$f_{\textrm{AM}}$ corresponds to the fundamental modulation frequency. The magnitude of the fundamental component is larger for square modulated displacement curves in all three modulation frequencies. On average the magnitude of the square modulated fundamental was 14.9\% higher compared to sine modulation.

However, the situation changes when looking at the higher harmonic frequencies. In Figure \ref{fig:square_vs_sine_160Hz}(b) it can be seen that the second harmonic component at 320 Hz has almost completely vanished for square modulation, but is still visible in the case of sine modulation. This can also be observed in Figure \ref{fig:DFT_560mVpp}(b) where the second harmonic component is at the noise level for square AM and significantly larger for sine AM. The same also applies to the 50 Hz modulation frequency in Figure \ref{fig:DFT_560mVpp}(a) which shows no existence of a second harmonic frequency. Only in the case of 1000 Hz modulation in Figure \ref{fig:DFT_560mVpp}(c) there seems to be a larger second harmonic component in the case of square modulation, but this is because the overall energy level at these high frequencies is already so low that the spectral noise contributes to the magnitude of these frequencies.

An opposite situation is observed in the case of 3$f_{\textrm{AM}}$ which is evident for square modulated displacement curves, but contains only a small amount of energy in the case of sine modulation. The effect is most pronounced at 50 and 160 Hz, but at 1000 Hz the energy contribution to third harmonic frequencies is due to the noise.

\subsection*{Simulations}

\begin{figure}[htbp!]
    \centering
    \subfigure[]
    {
        \includegraphics[height=5cm]{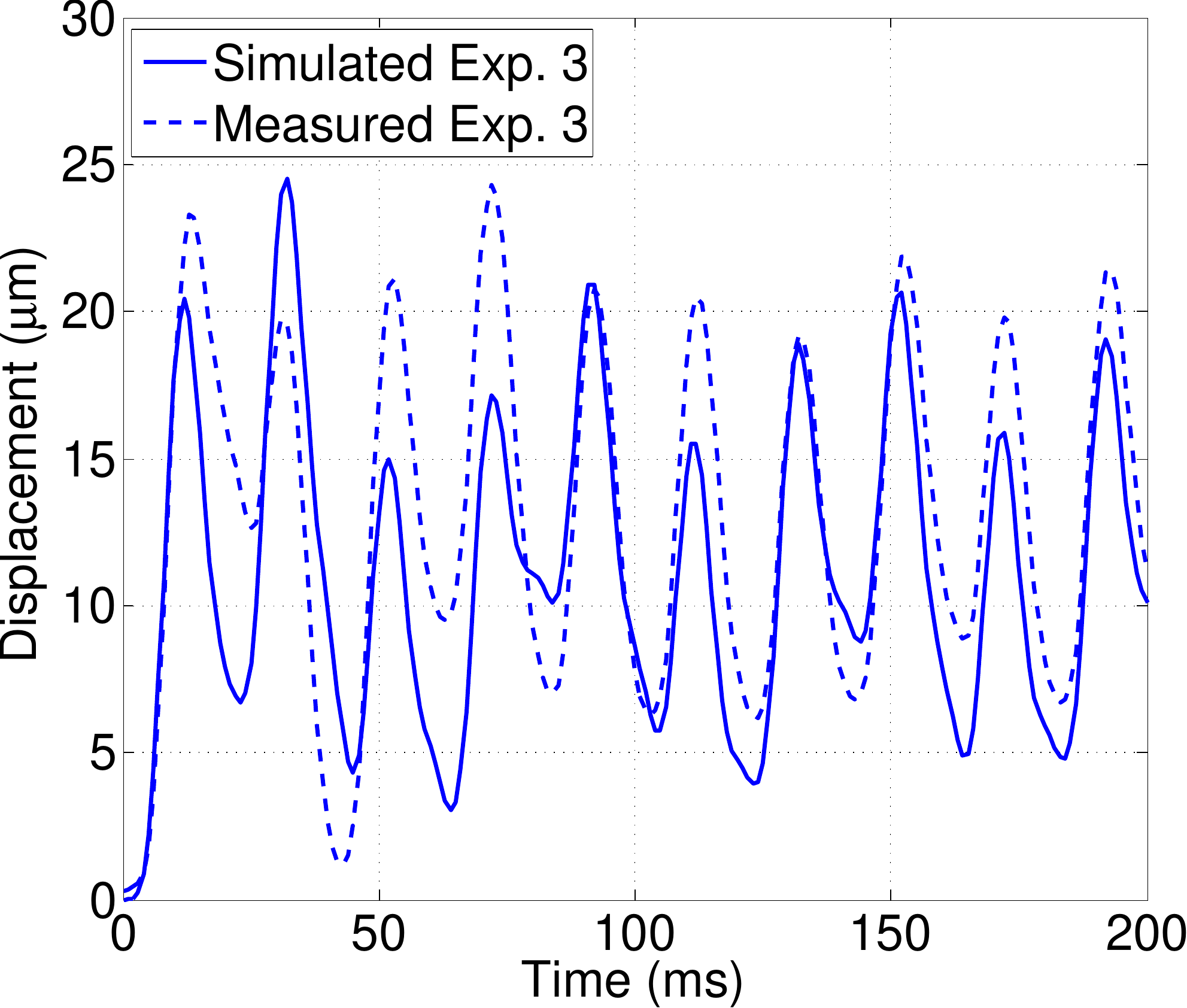}
    }
    \subfigure[]
    {
        \includegraphics[height=5cm]{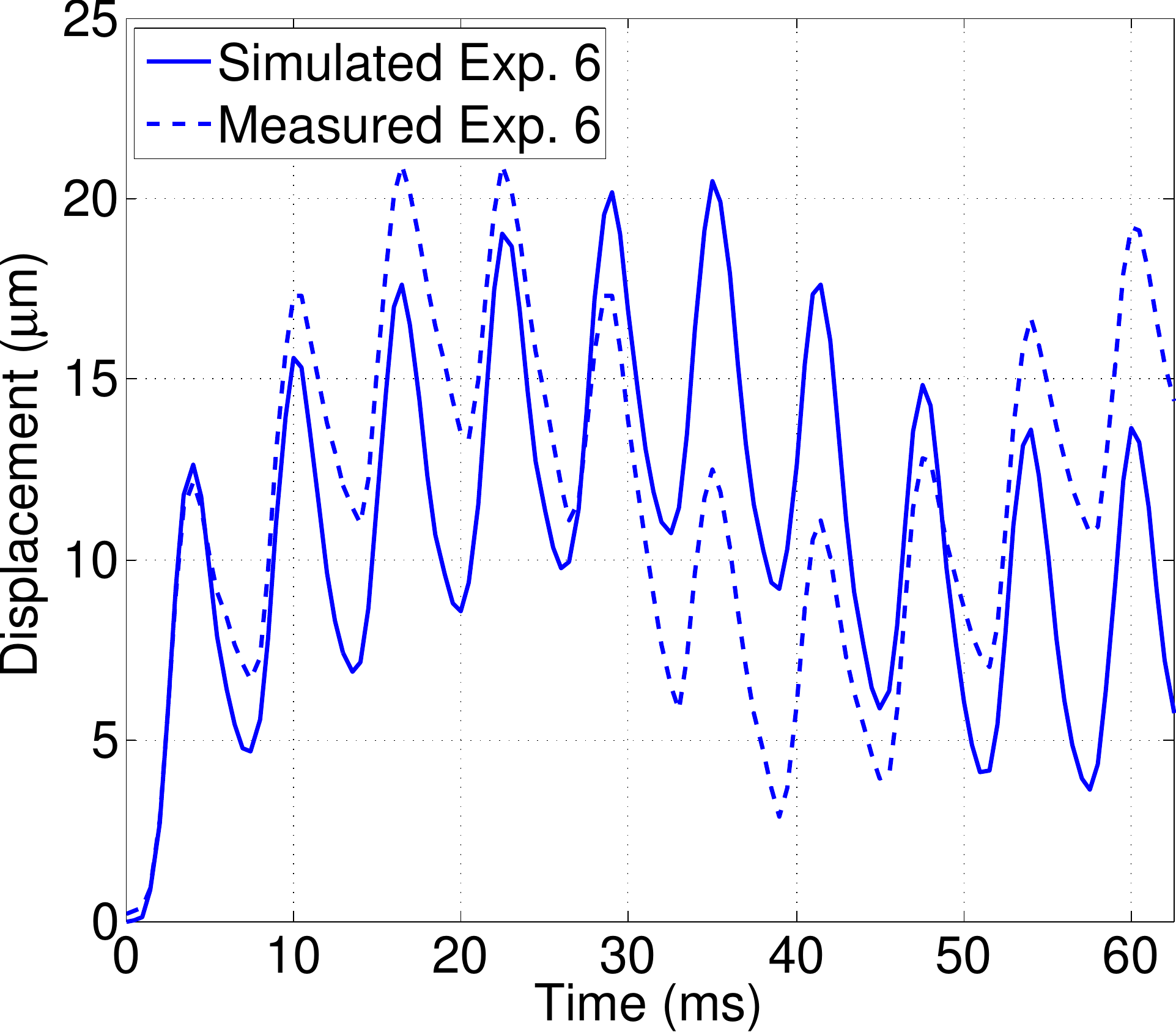}
    }
    \\
    \subfigure[]
    {
        \includegraphics[height=5cm]{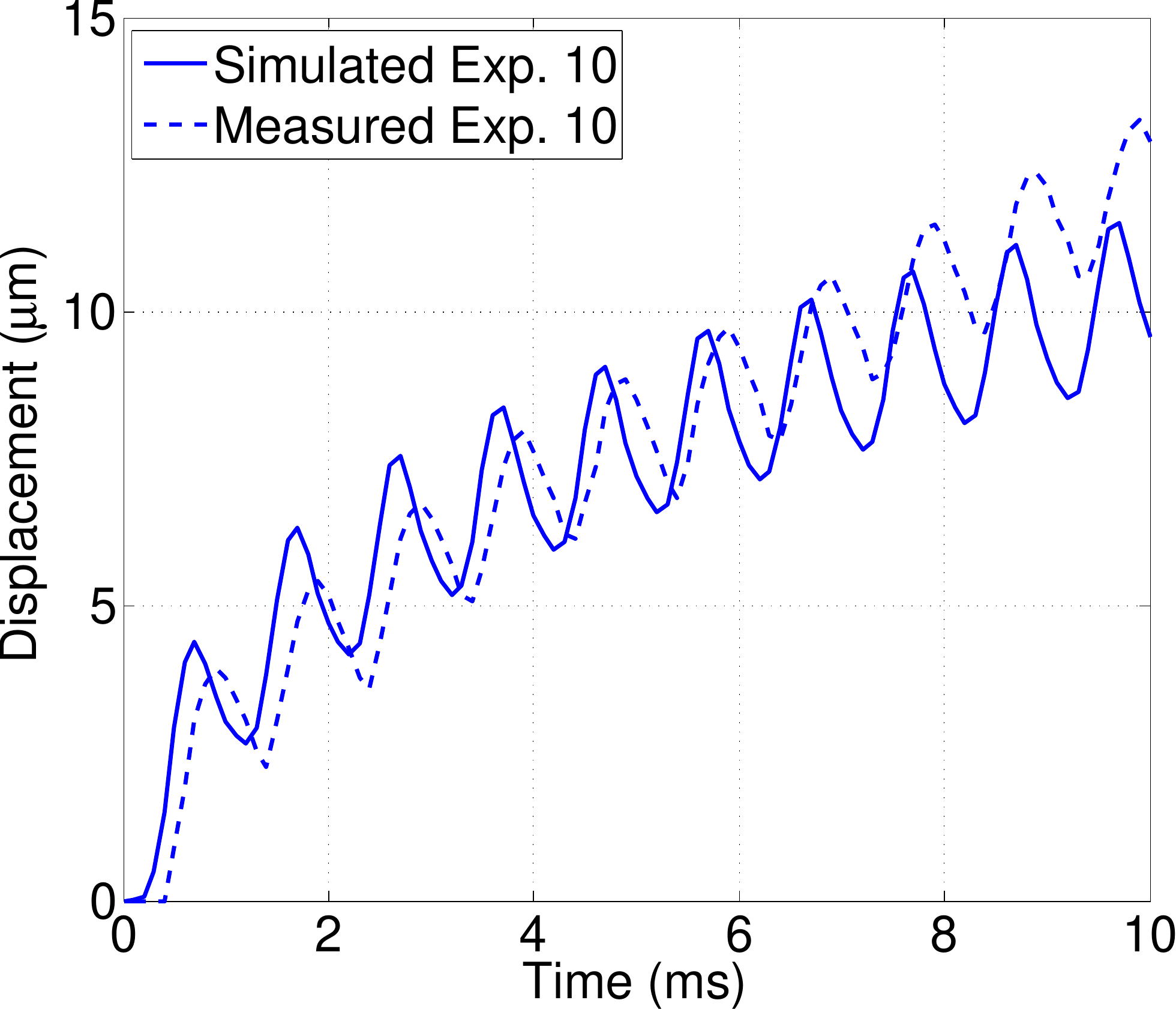}
    }
    \subfigure[]
    {
        \includegraphics[height=5cm]{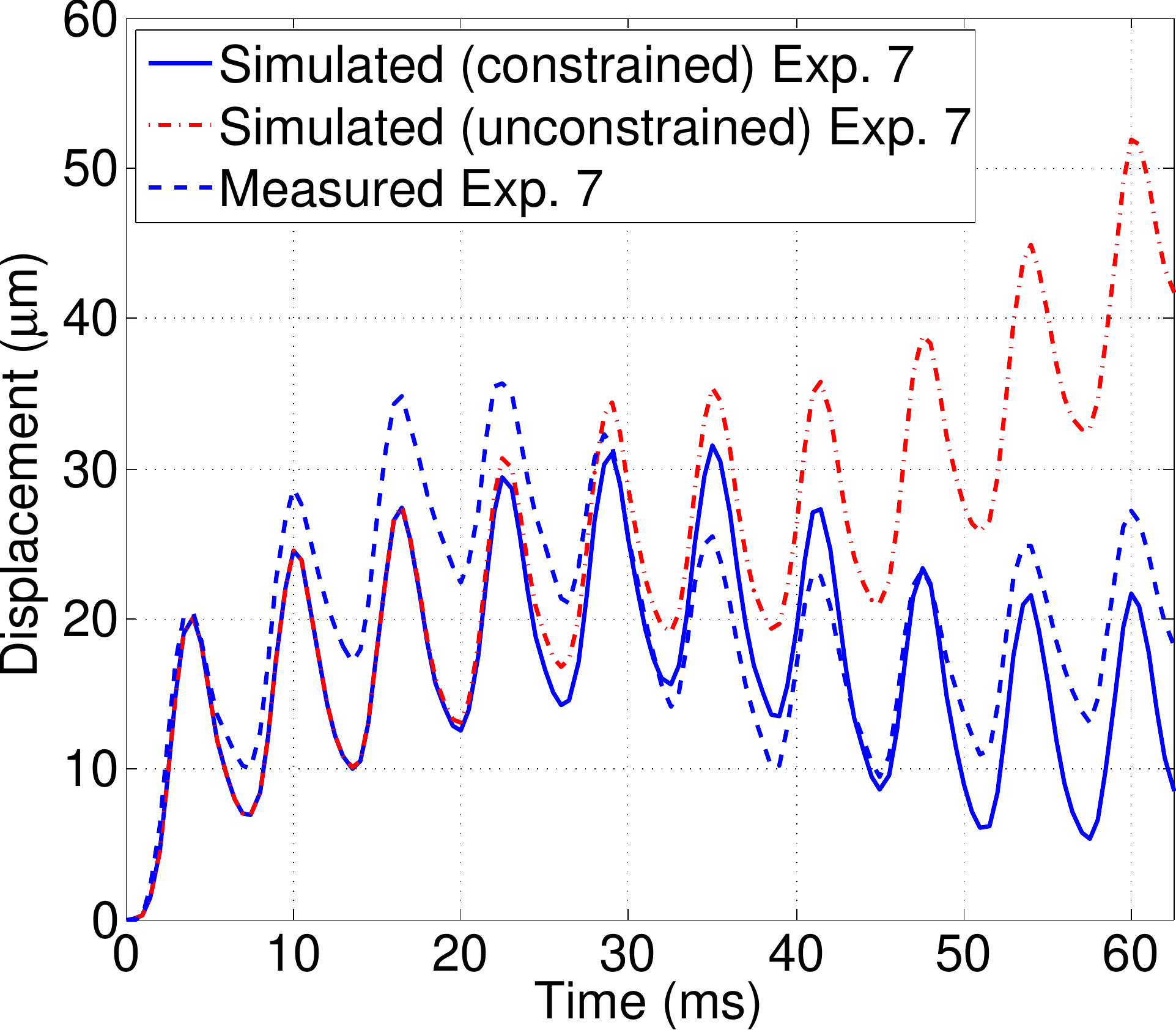}
    }
    \caption{Simulated displacement curves compared with the experimental data for (a) 50 Hz, (b) 160 Hz and (c) 1000 Hz sine amplitude modulation (AM). The effect of constrained vs. unconstrained boundary condition is shown in (d) together with the experimental data. The data corresponds to the experiments no. 3, 6, 10 and 7 in Table \ref{tab:ultrasound_parameters}, respectively.}
    \label{fig:simulation_amp_with_frequency}
\end{figure}

The simulated displacement curves together with the experimental data for 50, 160 and 1000 Hz sine AM are shown in Figures \ref{fig:simulation_amp_with_frequency}(a)-(c) respectively. All ten ARF excitation cycles are shown for each modulation frequency and the data are presented in absolute displacement values so that a quantitative comparison of the FEM model and the experiments can be carried out.

Figure \ref{fig:simulation_amp_with_frequency}(a) shows the simulated and measured displacement curves for 50 Hz sine AM. The simulation captures the dynamics that were observed experimentally. There is up to 5 $\mu$m discrepancy in the peak amplitude, but the simulation successfully reproduces the deeper relaxations of the measured curves at 40, 80, 120 and 160 ms which are associated with the low frequency baseline drift at around 25 Hz. Similarly, the baseline drift is clearly visible in Figure \ref{fig:simulation_amp_with_frequency}(b) which shows the data for 160 Hz sine modulation. The experimental data shows higher peak values during the first four cycles, but the low frequency baseline drift relaxes faster, and thus, ends up to lower values at around 40 ms. In both cases, the peak-to-peak amplitudes of the simulation data are slightly higher compared to the measured values. 

Figure \ref{fig:simulation_amp_with_frequency}(c) shows the data for 1000 Hz sine AM. The peak-to-peak displacement amplitudes of the simulation are similar to those of the experimental data, but the overall peak value after ten ARF cycles is slightly lower. Furthermore, a small phase shift of exists between the simulated and measured curves, whilst the peaks of the simulated data occur earlier. This is due to a delay in response of the phantom at the onset of ARF in the experiments, which causes 0.4 ms delay in the beginning of the displacement curve. Similar small delay was also observed in the case of 50 and 160 Hz sine modulation, but the phase shift is not clearly visible due to the longer period of the ARF cycles.

The effect of constrained boundary condition is demonstrated in Figure \ref{fig:simulation_amp_with_frequency}(d) which shows the simulated constrained and unconstrained displacement curves together with the corresponding experimental data with 160 Hz sine modulation. The unconstrained curve continues towards higher peak values whilst the constrained curve starts decreasing after four ARF cycles similarly to the experimental data, which causes the system to resonate at around 25 Hz.

\begin{figure}[t]
    \centering
    \subfigure[]
    {
        \includegraphics[height=6cm]{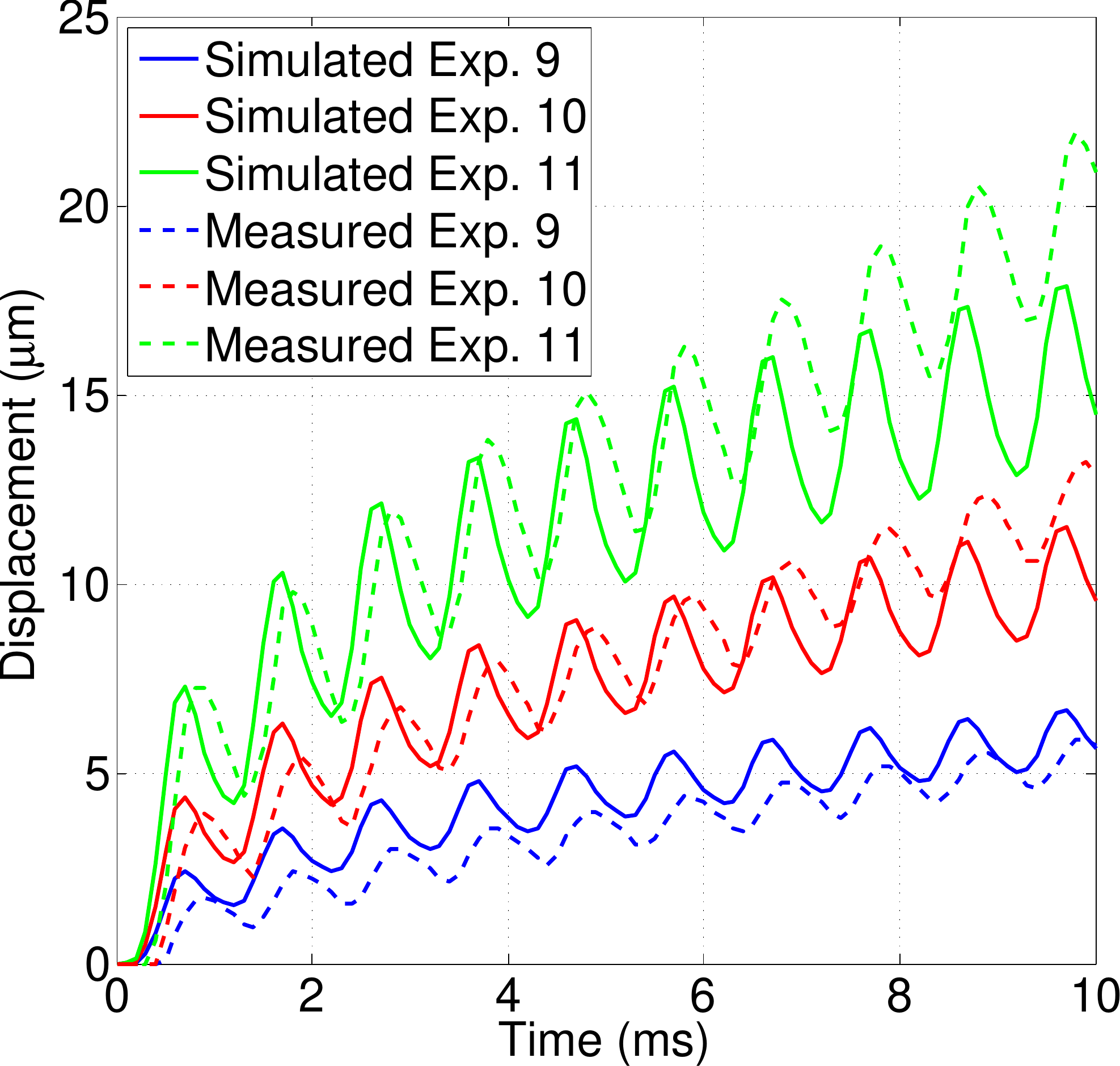}
    }
    \subfigure[]
    {
        \includegraphics[height=6cm]{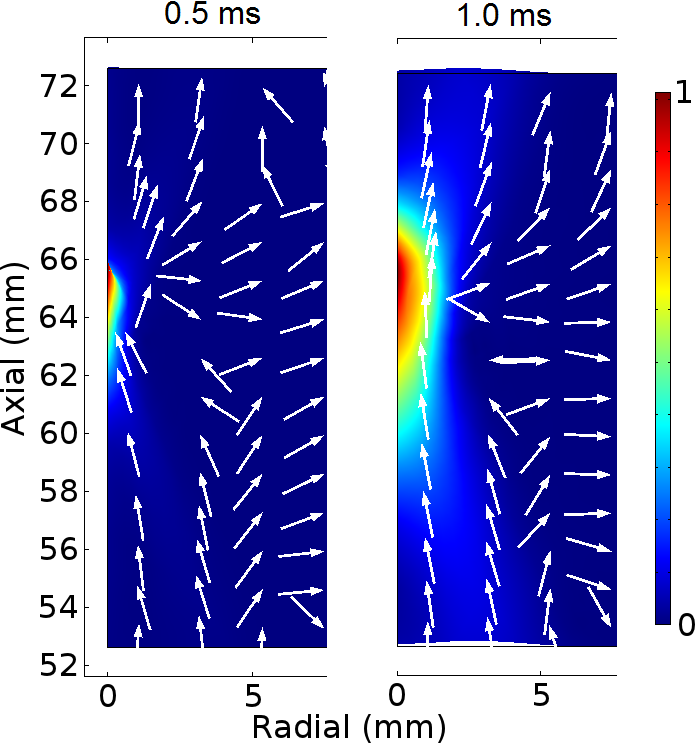}
    }
    \caption{Simulated displacement curves for (a) three different intensity levels with 1000 Hz sine amplitude modulation (AM) and (b) 2-D presentation of the phantom deformation field during the first cycle of the excitation. The white arrows represent the direction of deformation and the colour map shows the normalised magnitude of the displacement. The data corresponds to the experiments no. 9-11 in Table \ref{tab:ultrasound_parameters} in (a) and the experiment no. 11 in (b).}
    \label{fig:simulation_amp_with_power}
\end{figure}

Figure \ref{fig:simulation_amp_with_power}(a) shows the simulated displacement curves together with the corresponding experimental data using three different ultrasound intensity levels using sine AM at 1000 Hz. The simulated data achieves higher peak values at low intensity levels, but falls behind when the intensity of the ultrasound field is increased. Furthermore, the experimental data seems to grow slightly faster within ten ARF cycles whilst the growth of the simulated data is slower causing it to achieve lower peak amplitude in the end. The delay at the onset of ARF is also visible in the experimental data and is smaller when using higher intensity levels. For the three different ultrasound intensities of 1209, 2038 and 3013 W/cm$^2$ respective delays of 0.5, 0.4 and 0.3 ms are induced in the beginning of the displacement curve.

Figure \ref{fig:simulation_amp_with_power}(b) is shows a 2-D deformation of the phantom mesh during the first ARF cycle using 1000 Hz sine modulation. The white arrows represent the direction of the displacement at 0.5 and 1.0 ms during the ARF excitation cycle and the colour map shows the normalised magnitude of the displacement. At 0.5 ms the magnitude of the ARF is the highest and the material in front of the focal point is moved towards the ultrasound focal point both in axial and lateral directions. Simultaneously, due to incompressibility, the material beyond the focal point is pushed away from the focal point towards the edge of the phantom causing it to deform. After a full ARF cycle at 1.0 ms the material in the focal area has moved in the axial direction whilst the front and rear edge of the phantom have slightly deformed. This deformation causes the upward trend of the displacement curves and the static component in the frequency spectrum.

\subsection*{Error estimation}

The main sources of experimental error were due to the registration of the optical and acoustic focal points and the alignment of the microsphere in the FOV of the objective. The determination of the acoustic focal point in lateral and elevation directions was possible within the diameter of the active element of the hydrophone $\pm$75 $\mu$m, which was accurate compared to the full width at half maximum (FWHM) of the transducer focal point. In the axial direction the accuracy was determined by the step size of the positioning system $\pm$1 $\mu$m.

The minimum distance between ultrasound focal point and the phantom edge was 3.3 mm which was due to the working distance of the objective. \citet{bouchard2009optical} reported the effect of shear wave reflection on the displacement curves from the phantom edge using single excitation pulses. Similar phenomenon was observed in the experiments when a single excitation pulse was used resulting in a small peak of approximately 1 $\mu$m after the main excitation (less than 5\% of the main peak). However, when harmonic excitation pulses were used, the shear wave refection was not visible in the displacement curves. Furthermore, no differences in the displacement curves were observed when comparing the simulations with finite and infinite domains.

The first main source of error arose from the alignment of the transducer and objective focal points. First of all, the physical tip diameter of the hydrophone was 300 $\mu$m, which was approximately four times larger than the FOV of the objective (80 $\mu$m). This affected the alignment accuracy in the lateral direction, because the tip of the needle hydrophone could not completely fit in the lateral FOV of the objective. Secondly, the alignment in the elevation direction was determined by finding the visually sharpest point of the needle hydrophone tip. This could result in a maximum alignment offset of $\pm$110 $\mu$m, which is still within the FWHM of the acoustic focal point. The alignment accuracy in the axial direction was approximately $\pm$10 $\mu$m based on the accuracy of which the tip of the hydrophone could be positioned in the middle of the FOV.

The second source of error was a result of positioning the phantom in the optical focal point of the objective. Although the magnification of the objective was considerably higher than used in the earlier optical tracking studies \citep{bouchard2009superficial, bouchard2009optical, czernuszewicz2013experimental}, positioning the microsphere in the elevation direction could only be done by finding the visually sharpest point in the image. However, taking into account the small size of the microsphere (10.23 $\mu$m), this error is significantly smaller than what resulted from the registration of optical and acoustic focal points.

The shift of the maximum ARF in axial direction due to refraction effects in the phantom was estimated with nonlinear ultrasound simulations. The optical focal point of the objective was set at the maximum peak-positive pressure. The peak-positive pressure shift was found to be approximately 0.6 mm in axial direction away from the transducer. However, the location of the peak-positive pressure is slightly different from that of the peak intensity value and hence the maximum ARF. In simulations the peak intensity was found to occur 0.2 mm before the peak-positive pressure, and therefore, combining these two effects result in 0.4 mm shift of the maximum ARF from the optical focal point in axial direction. The total focal shift is less than 9\% of the FWHM pressure of the focal point along the same axis (4.5 mm).

The analysis of the image data could also result in errors at a sub-micrometre scale. The correlation coefficient value drops during the excitation phase, which affects the registration accuracy of the two subsequent image frames. The correlation coefficient values were typically high over 0.8, but in the case of high modulation frequencies the accuracy might drop slightly due to the exposure time of the high-speed camera sensor. This error could be reduced using higher frame rates with a more powerful light source.

The accuracy of the simulation model to mimic the dynamics of the experiments is affected by several factors. The first source of simulation error comes from the location of the ultrasound focal point with respect to the marker. As stated earlier, the location of the microsphere in lateral and elevation directions is subject to a small misalignment in the experiment whereas in the simulation the marker is placed exactly in the centre of the geometric focal point. The second source of error arises from determining the mechanical parameters of the phantom material. Whilst attenuation, density, speed of sound and Young's modulus values could be accurately measured, the choice of viscoelastic model and its parameters might not fully replicate the real dynamics of the medium. For example, the choice of Poisson's ratio has been shown to have an effect on the axial displacement amplitude \citep{palmeri2005finite}. Adjusting material properties to improve agreement was not employed here. 

The third source of error is related to the simulation of the nonlinear ultrasound field. Although the intensity of the ultrasound field could be accurately matched with the experimentally measured values, the size of the simulated focal point was observed to be slightly smaller. Smaller focal point, and consequently smaller displacement volume, has been shown to lead to shorter relaxation time of the material \citep{czernuszewicz2013experimental}. Finally, the phantom geometry in the FEM model was chosen to be axisymmetric cylinder due to the reduced computational time, which does not affect the displacement estimation accuracy in micrometre scale, but could have an effect if larger deformations are induced.

\section*{Discussion}

Characterisation of harmonic displacements using different levels of ARF, AM waveforms and modulation frequencies was conducted in a tissue-mimicking phantom. The shapes of the harmonic displacement curves in the time domain correspond to the optically measured harmonic excitations by \citet{bouchard2009superficial}. Although they did not characterise the displacement amplitudes with respect to ARF, the magnitudes of the peak-to-peak displacements at the modulation frequency of 100 Hz are of the same order of magnitude at around 15 $\mu$m. \citet{curiel2009vivo} measured the average harmonic displacement amplitudes in \textit{in vivo} rabbit muscle. Given the parameters of their experiments and assuming the mean sound speed in soft tissue (1540 m/s), the levels of ARF varied between 0.14 and 0.29 N/cm$^{3}$ resulting in average displacement amplitudes of 19-43 $\mu$m using 75 Hz modulation frequency. However, the harmonic amplitude measurements were performed during HIFU ablation, and hence, cannot be related to a specific stiffness value of the tissue \citep{sapin2010temperature, benech2010monitoring}. Furthermore, ARF is dependent on the attenuation of the tissue, which changes considerably during ablation \citep{jackson2014nonlinear}.

One of the observations in the measurement results was that larger peak-to-peak displacement amplitudes were achieved using square modulation: on average sine modulated displacements curves had 19.5\% lower peak-to-peak amplitude values compared to square modulation. This is because the time average ARF of sine AM waveform within one cycle is 25\% lower than that of square modulation, which results in smaller energy deposition with time. Furthermore, the instantaneous rise of ARF during the first half of the excitation cycle in square modulation causes greater displacement amplitude at the beginning. During the second half ARF is completely turned off, which results in a steeper slope of the displacement curve and consequently lower values.

The average peak-to-peak displacements were found to increase with simulated ARF. The dependence was approximately linear up to 10 $\mu$m after which the slope decreased slightly. This is consistent with \citet{zaitsev2004focused} who studied how different acoustic output power affected the displacement amplitude in a silicone phantom using 10 ms ultrasound bursts and found the dependence to be approximately linear. However, this dependence is affected by the viscoelastic properties of the material and the nonlinearity of the ultrasound field.

The dependence of the displacement amplitude on modulation frequency was observed to be nonlinear. Similar results were also shown in experiments by \citet{curiel2009vivo} and later in simulations by \citet{heikkila2010local}, who examined how different modulation frequencies affect normalised harmonic displacement amplitudes. In both studies the harmonic displacement amplitude was found to decrease with modulation frequency, whilst largest displacement amplitudes were achieved using low modulation frequencies. Although this relation has been a decreasing power fit in the previous cases, the shape of this nonlinear dependence is related to the frequency response of the material \citep{liu2007viscoelastic}.

Another property which was observed to change with modulation frequency was the phase shift between the displacement curves and the applied ARF. For both sine and square modulation, the absolute phase shift was shown to increase with the modulation frequency. However, the magnitude of the phase shifts was not found to vary with different levels of ARF when the modulation frequency was kept constant. Similar phase shifts were also observed in experiments by \citet{maleke2006single}, who generated harmonic sine AM displacements in gelatin phantoms. They found the magnitudes of the phase shifts to decrease with increasing phantom stiffness using a constant modulation frequency of 50 Hz. \citet{bouchard2009superficial} used optical tracking method to measure the phase shifts in a gelatin phantom. They used sine AM waveforms with modulation frequencies from 50 to 300 Hz and found the absolute phase shifts to increase within this range.

The phase shifts occur due to the viscoelastic properties of the medium, and therefore, are dependent on the phase response of the material \citep{liu2007viscoelastic, vappou2009quantitative}. If the material was completely elastic, it would behave as a linear spring where all the deposited energy in the excitation phase would be deposited in storage modulus and released in phase with the ARF. However, because the tissue-mimicking phantom used in the experiments is viscoelastic material, part of the energy is lost to viscosity. Therefore, when a periodic force is applied to a material which is not perfectly elastic, the strain is not in phase with the force.

In all cases a direct current (DC) offset of displacement was observed with time. The static component of the displacement curve was also observed in the phantom experiments conducted by \citet{curiel2011localized}. They used 75 Hz modulation frequency to transmit five cycles of harmonic ultrasound pulses to three different locations in a phantom and rabbit thigh. The frequency spectrum analysis of the obtained displacement curves showed a strong static component at 0 Hz, but the time window used was not long enough to capture the baseline drift. The low frequency baseline drift was observed in simulations by \citet{konofagou2003localized} and \citet{heikkila2010local}, who postulated them to be due to non-zero mean of the ARF. The non-zero mean could explain the existence of the static component, but in addition to the DC offset a low frequency oscillation at around 25 Hz was observed for long enough time windows. This is more likely related to the resonant frequency of the medium and its surroundings. This behaviour became evident in the simulation model, where the constrained outer surface of the phantom caused a low frequency baseline drift to the displacement curves. Therefore, the baseline drift in the experimental data is caused by fixing the phantom to its holder from the both ends in the lateral direction.

The characterisation of harmonic frequencies showed differences between sine and square modulated displacement curves. The magnitude of the fundamental component 1$f_{\textrm{AM}}$ was found to be consistently greater for square modulation, which was a direct result of larger average peak-to-peak displacement amplitudes. Another difference was found in the ratio between the components at 2$f_{\textrm{AM}}$ and 3$f_{\textrm{AM}}$. Where square modulated displacement curves had barely no energy at 2$f_{\textrm{AM}}$, sine modulation showed a detectable peak in the frequency spectrum. In contrast, the magnitude of 3$f_{\textrm{AM}}$ was noticeably larger for square modulation as opposed to sine modulation. The explanation for the absence of a second harmonic component in the case of square modulation is in the fundamental property of the Fourier expansion of a square wave, which contains only components of odd-integer harmonic frequencies. In a theoretical situation the magnitude of odd-integer harmonic frequencies would be zero, but in real world situation some energy still remains.

The simulation model successfully replicated the dynamics of the experimental data inducing some of the lower frequency oscillations. The small delays in the beginning of the displacement curves using sine AM were due to the behaviour of ARF according to Eq. (\ref{eq:sine_modulation}) where ARF is modulated using normalised $P_{\textrm{sine}}^{2}$. This causes the ARF to increase slowly in the beginning causing smaller slope. The overall peak amplitudes of the simulations were up to 5 $\mu$m lower compared to the measured values, whereas the peak-to-peak amplitudes showed slightly larger values. This is most likely due to the chosen viscoelastic model and its parameters, which could not completely mimic the relaxation process of the medium with different modulation frequencies. For example, the peak-to-peak displacement amplitudes were higher in the simulations at 160 Hz compared to the experimental data, whilst at 1000 Hz the amplitudes were approximately the same, which suggest that the chosen parameters were better suited for high frequencies. Therefore, a more generalised viscoelastic model which takes into account multiple relaxation times, such as Maxwell-Wiechert model, could be more precise in predicting the relaxation process of the medium when different modulation frequencies are used. However, this requires accurate determination of the parameters for the viscoelastic model, which was outside the scope of this study.

\section*{Conclusions}

No previous work on characterisation of displacement amplitudes and phase-shifts with respect to ARF has been reported. Although previous reports have shown single pulse displacement amplitudes with respect to ultrasound field intensity, the frequency dependent attenuation properties of tissue are significant when nonlinear ultrasound fields are used. Here is shown that measured displacements are in quantitative agreement with simulated displacements when the tissue properties and ultrasound fields are known. This is useful in diagnostic applications such as ARFI imaging and HMI.

Comparison of square and sine amplitude-modulation is useful in determining suitable modulation waveforms for diagnostic applications such as HMI and EMA at the frequency appropriate for the given modality. The characterisation of frequency components of different amplitude-modulation waveforms is useful in diagnostic applications such as EMA imaging which rely on quantifying the frequency spectrum components. The comparison of square and sine modulation frequency components helps to define suitable waveforms to achieve maximum fundamental frequency component with minimum harmonic frequency components.

Optical tracking of harmonic displacements in a tissue-mimicking phantoms was successfully conducted. The square modulated ARF waveforms yielded higher average peak-to-peak amplitude values when compared to sine modulation. Furthermore, the displacement amplitudes showed increasing linear dependence on ARF up to about 10 $\mu$m after which the slope decreased with both sine and square modulation. Phase shifts between the ARF and AM displacement curves were also observed. The magnitude of the phase shift was approximately independent of ARF but increased with the modulation frequency in both cases.

In the frequency domain, energy was observed at the fundamental modulation frequency, its harmonics and around 0 Hz. The magnitude of the fundamental modulation frequency was consistently higher in the case of square modulation, whereas the ratio between the second and third harmonic components differed as opposed to sine modulation. The second harmonic was almost non-existent in the case of square modulation, but energy was found in the third harmonic. An opposite situation was observed in the case of sine modulation, where a stronger second harmonic and a very low energy third harmonic were detected.

An advance over other studies is that a quantitative comparison of simulated and measured displacements was conducted. The simulation results were found to follow the dynamics of the experimental data, but small differences were seen in absolute peak and peak-to-peak displacement amplitudes. This was most likely due to the chosen viscoelastic model, which was unable to account for multiple modulation frequencies. A choice of more general viscoelastic model might be able to predict the dynamics more accurately when different modulation frequencies are used.

These results show that FEM models can quantitatively capture ARF induced displacements over a range of amplitudes and modulation frequencies. This provides a numerical tool to determine ultrasound intensity levels, AM waveforms and modulation frequencies in diagnostic applications, which are based on generating and tracking harmonic displacements in tissue.
        
\section*{Acknowledgements}

V.~S. acknowledges the support of the RCUK Digital Economy Programme grant number EP/G036861/1 (Oxford Centre for Doctoral Training in Healthcare Innovation) as well as the support of Instrumentarium Science Foundation, Jenny and Antti Wihuri Foundation and Finnish Cultural Foundation. R.~C. and D.~E. acknowledge the support of EPSRC grant number EP/K02020X/1. 





\pagebreak

\bibliographystyle{UMB-elsarticle-harvbib}
\bibliography{arXiv_optical_tracking}

\end{document}